\documentclass[twocolumn,superscriptaddress,preprintnumbers,amsmath,amssymb,prc,nofootinbib]{revtex4-2}

\usepackage{graphicx}
\usepackage{dcolumn}
\usepackage{bm}
\usepackage{ifpdf}
\usepackage{epstopdf}
\usepackage{slashed}
\usepackage{amsfonts}
\usepackage{mathrsfs}
\usepackage{amssymb}
\usepackage{multirow}
\usepackage{CJK}
\usepackage{xcolor}
\usepackage{ulem}

\def\lesssim{\ \raise.3ex\hbox{$<$}\kern-0.8em\lower.7ex\hbox{$\sim$}\ }
\def\gesim{\ \raise.3ex\hbox{$>$}\kern-0.8em\lower.7ex\hbox{$\sim$}\ }

\begin{document}
\begin{CJK}{UTF8}{ipxm}
\preprint{RIKEN-iTHEMS-Report-23}

\title{Medium-induced bosonic clusters in a Bose-Fermi mixture: 
\\
{Towards simulating cluster formations in neutron-rich matter}}

\author{Yixin Guo (郭一昕)}
\email{guoyixin1997@g.ecc.u-tokyo.ac.jp}
\affiliation{Department of Physics, Graduate School of Science, The University of Tokyo, Tokyo 113-0033, Japan}
\affiliation{RIKEN Interdisciplinary Theoretical and Mathematical Sciences Program (iTHEMS), Wako 351-0198, Japan}

\author{Hiroyuki Tajima (田島裕之)}
\email{hiroyuki.tajima@tnp.phys.s.u-tokyo.ac.jp}
\affiliation{Department of Physics, Graduate School of Science, The University of Tokyo, Tokyo 113-0033, Japan}

\date{\today}

\begin{abstract}
Considering bosonic atoms immersed in a dilute Fermi gas, we theoretically investigate medium-induced bosonic clusters associated with fermion-mediated two- and three-body interactions. 
Using the variational approach combined with the fermion-mediated interactions, we numerically calculate the binding energies of two- and three-body bosonic clusters in a one-dimensional system.
It is found that the bosonic clusters can be formed even with a repulsive boson-boson interaction due to the fermion-mediated interactions.
Our results would be relevant for ultracold-atomic systems as well as analog quantum simulations of alpha clusters in neutron-rich matter.
\end{abstract}

\maketitle

\section{Introduction}\label{sec:I}

Many-body physics, which we encounter in various contexts such as condensed-matter and nuclear physics,
has been presenting challenging problems for a long time in modern physics.
Strong interactions and multiple degrees of freedom lead to nontrivial phenomena including many-body bound states~\cite{Freer2018Rev.Mod.Phys.90.035004} and superfluidity or superconductivity~\cite{Zwierlein2006Science311.492--496,Balents2020Nat.Phys.16.725--733}.

In nuclear physics, strong nuclear forces facilitate the formation of light clusters such as $\alpha$ particles, which are considered as a potential stable subunit within nuclei.
Previous studies showed that $\alpha$ clusters can appear across the table of nuclides from light- or medium-mass elements to heavy or even superheavy elements~\cite{Horiuchi2012Prog.Theor.Phys.Suppl.192.1--238, Freer2018Rev.Mod.Phys.90.035004,Bai2019Phys.Rev.C99.034305}. 
The $\alpha$ clustering has provided great assistance in describing various nuclear properties with its specific structure and properties.
In neutron-rich nuclei, the extra (valence) neutrons may be exchanged between $\alpha$-cluster cores~\cite{Itagaki2000Phys.Rev.C61.044306,Elhatisari2017Phys.Rev.Lett.119.222505}.
Interestingly, recent experiments have indicated the formation of $\alpha$ clusters in the surface region of heavy neutron-rich nuclei~\cite{Tanaka2021Science.371.260}, which can be regarded as the in-medium clusters in neutron-rich matter~\cite{Moriya2021Phys.Rev.C104.065801}. 
The formation of the medium-induced bound states in neutron-rich matter has also been discussed in connection with diprotons~\cite{tajima2023polaronic}.
However, it is still challenging to see how impuritylike particles behave and form bound clusters in the presence of a background Fermi sea.

Nevertheless, this problem can be tackled in an analogous system, that is, ultracold-atomic systems~\cite{Chevy2010Rep.Prog.Phys.73.112401,Schmidt2018Rep.Prog.Phys.81.024401}.
A controllable ultracold-atomic gas, where unique experimental access to tunable interactions is available, is nowadays one of the best candidates to investigate the unconventional many-body states in a systematic way. 
Recently, a quasiparticle picture of impurities immersed in background media, called a polaron~\cite{Landau1933Phys.Z.Sowjetunion3.664,landau1948effective}, has been extensively examined in cold-atomic systems~\cite{Chevy2010Rep.Prog.Phys.73.112401,Massignan2014Rep.Prog.Phys.77.034401,Schmidt2018Rep.Prog.Phys.81.024401}.  
Such polaronic states in the medium consisting of fermionic (bosonic) atoms are referred to as Fermi (Bose) polarons,
which provide us with opportunities to obtain a fundamental understanding of the role of many-body correlations as well as medium properties in a quantitative manner.
As a step further, such an investigation can also be extended to many-polaron systems.
Because of the medium-induced interactions,
polarons may feel the effective attraction leading to multibody bound polaronic states called
bipolarons~\cite{Fisher1989J.Phys.:Condens.Matter1.5567,Alexandrov1994Rep.Prog.Phys.57.1197} and tripolarons~\cite{Rice1986Phys.Rev.B34.4139--4149,King2015Phys.Rev.B91.024412}.
For instance, the bound bipolaron states formed by repulsive phonon-mediated interactions~\cite{Sous2017Phys.Rev.A96.063619} and Bose bipolarons with an induced nonlocal interaction mediated by density oscillations~\cite{Camacho-Guardian2018Phys.Rev.Lett.121.013401} have been theoretically studied.

While it is known that ultracold atoms are unstable with respect to the formation of three-body bound states because of the three-body recombination~\cite{Eismann2016Phys.Rev.X6.021025},
few-body properties such as Efimov effects can be investigated in this system via the three-body loss measurement~\cite{Barontini2009Phys.Rev.Lett.103.043201}.
In this regard, it is worth investigating two- and three-body physics in ultracold-atomic gases.
In such a direction, the in-medium effects on Efimov trimers and medium-induced three-body states are of interest in atomic systems~\cite{Niemann2012Phys.Rev.A86.013628,nygaard2014efimov,Kirk2017Phys.Rev.A96.053614,Sun2019PhysRevA.99.060701,tajima2019quantum,Tajima2021Phys.Rev.A104.053328,Tajima2022Phys.Rev.Research4.L012021,Guo2022Phys.Rev.A106.043310,Musolino2022Phys.Rev.Lett.128.020401,Guo2023Phys.Rev.B107.024511}.
While in bosonic~\cite{Burt1997Phys.Rev.Lett.79.337--340, Haller2011Phys.Rev.Lett.107.230404} and fermionic~\cite{Ottenstein2008Phys.Rev.Lett.101.203202,Huckans2009Phys.Rev.Lett.102.165302} systems, the three-body recombinations involving identical particles are enhanced and suppressed, respectively, due to different quantum statistics, it remains elusive how three-body processes are modified in the many-body background. 
The effective interpolaron interaction mediated by the degenerate Fermi gas further complicates the problem~\cite{PhysRevA.79.013629,PhysRevA.102.063321,Bougas2021New.J.Phys.23.093022}.
Recently, the fermion-mediated interaction~\cite{DeSalvo2019Nature568.61--64,PhysRevLett.124.163401,baroni2023mediated}, which is analogous to the Ruderman-Kittel-Kasuya-Yosida (RKKY) interaction~\cite{Ruderman1954Phys.Rev.96.99--102,Kasuya1956Prog.Theor.Phys.16.45--57, Yosida1957Phys.Rev.106.893--898}, and the suppression of unitary three-body loss in a degenerate Bose-Fermi mixture with effects of both the Fermi statistics and the RKKY interaction~\cite{Chen2022Phys.Rev.Lett.128.153401} have been experimentally studied.

In this paper, we theoretically investigate the fate of bosonic atoms immersed in a dilute degenerate Fermi gas.
Using the fermion-mediated two- and three-body interactions at the low-energy and long-wavelength limit, 
we discuss how in-medium bosonic clusters are formed in the presence of background Fermi sea with arbitrary boson-boson and fermion-boson interactions.
From the experimental viewpoint, these in-medium few-body bosonic clusters can be associated with the three-body recombination process~\cite{Naidon2017Rep.Prog.Phys.80.056001}. Although we consider a one-dimensional system for simplicity, our setup is similar to the previous work for $\alpha$ particles in cold neutron matter in three dimensions~\cite{Moriya2021Phys.Rev.C104.065801}. In this regard, our study proposes a way to perform the quantum simulation of $\alpha$ clustering in neutron-rich matter by using an ultracold-atomic system. 

This paper is organized as follows.
The theoretical framework is presented in Sec.~\ref{sec:II}, where we show the Hamiltonian for a one-dimensional system with both bosonic and fermion-mediated interactions. 
We apply a variational method for in-medium three-body states on top of the Fermi sea to this model. 
In Sec.~\ref{sec:III}, we first discuss a special case with zero fermion-mediated interaction, i.e., bare bosonic potential in vacuum.
And then we show our numerical results for the bound states in the fermion medium.
Finally, a summary and perspectives will be given in Sec.~\ref{sec:IV}.
In the following, we take $\hbar=c=k_{\rm B}=1$.
The system size is taken to be a unit.

\section{Theoretical framework}\label{sec:II}

\begin{figure}[t]
  \includegraphics[width=0.5\textwidth]{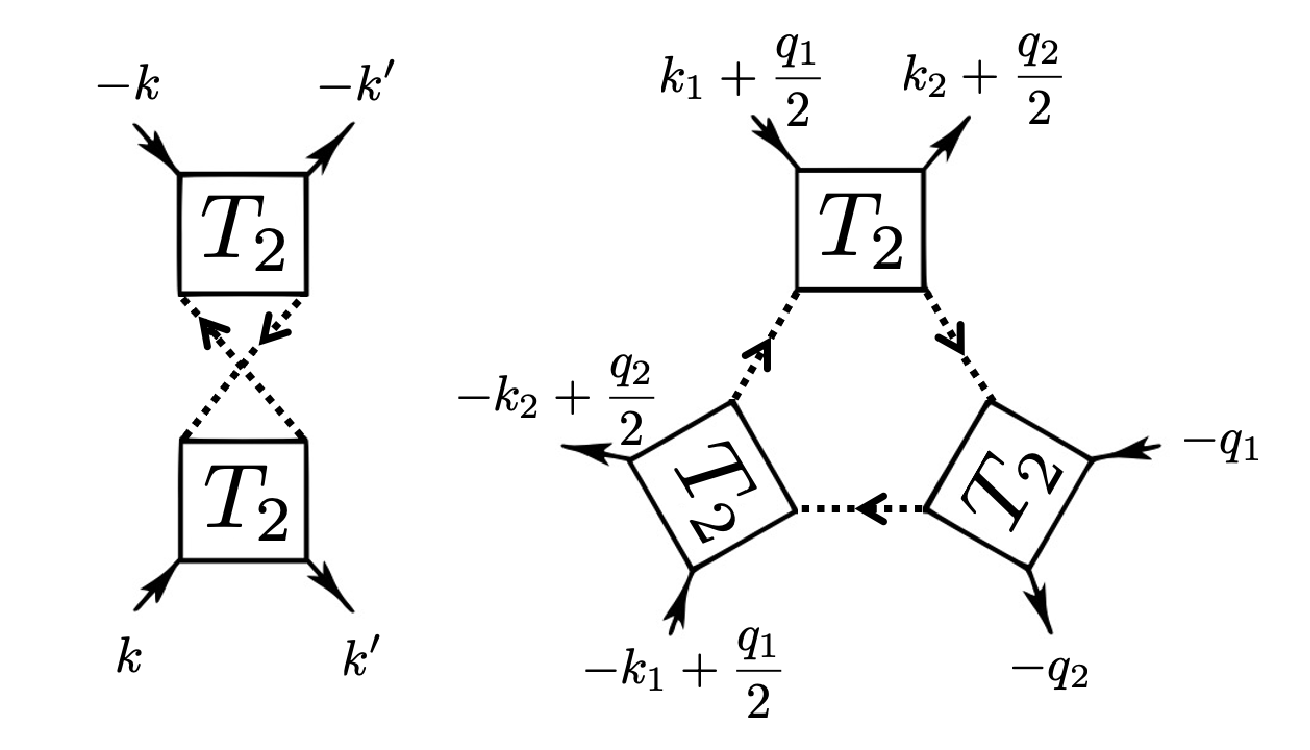}
  \caption{
  {
  Feynman diagrams for the medium-induced two- and three-body interactions~\cite{Moriya2021Phys.Rev.C104.065801,Tajima2021Atoms9.18}.
  For the medium-induced two-body interaction, the outgoing (incoming) momenta of bosons are given by $k'$ and $-k'$ ($k$ and $-k$), while they are $k_2+q_2/2$, $-k_2+q_2/2$, and $-q_2$ ($k_1+q_1/2$, $-k_1+q_1/2$, and $-q_1$) for the three-body interaction. 
  The internal dotted lines denote the Green's function of a medium fermion, and the arrows represent the direction of momentum and energy transfer in each propagator.}
  }\label{fig:digram}
\end{figure}

Here we consider the one-dimensional system of bosons with bare mass $m_b$ in a dilute gas of fermions with bare mass $m_f$ at zero temperature ($T\rightarrow 0$).
We start with diagrammatic derivation of the medium-induced two- and three-body interactions among bosons in a Fermi sea. 
{As depicted diagrammatically in Fig.~\ref{fig:digram}, the induced two-body~\cite{Patton2011Phys.Rev.A83.051607} and three-body~\cite{Tajima2021Atoms9.18} interactions can be obtained via the two-body $T$ matrix, $T_2(q,\omega)$~\cite{Tajima2021Atoms9.18}, for boson-fermion coupling, which just corresponds to the summation of the ladder diagram.}
{
\begin{widetext}
In detail, the results up to the leading order are given as
\begin{align}
V_{\mathrm{eff}}^{(2)}\left({q}, i v_{\ell}\right)
=
 T \sum_{{p}, \omega_n}
\left[T_2(k+p,\omega_k+\omega_n)
T_2(p+q-k,\omega_n+ v_{\ell}+\omega_{-k})\right] 
G\left({p}+{q}, i \omega_n+i v_{\ell}\right) 
 G\left({p}, i \omega_n\right),
\end{align}
and
\begin{align}
V_{\mathrm{eff}}^{(3)}\left({k}, {q}, i v_{\ell}, i v_u\right)=\,& 2 
T \sum_{{p}, \omega_n}
\left[
T_2(-k_1+q_1/2+p+k-q/2,\omega_{-k_1+q_1/2}+\omega_n+v_{\ell}-v_u)
\right.\nonumber\\&\left.\times
T_2(-q_1+p+k+q/2,\omega_{-q_1}+\omega_n+v_{\ell})
T_2(k_1+q_1/2+p,\omega_{k_1+q_1/2}+\omega_n)
\right]
\nonumber\\
&\times
G\left({p}, i \omega_n\right) 
 G\left({p}+{k}+{q} / 2, i \omega_n+i v_{\ell}\right) 
G\left({p}+{k}-{q} / 2, i \omega_n+i v_{\ell}-i v_u\right),
\end{align}
respectively.
\end{widetext}
Here, the two-body $T$ matrix $T_2(q,\omega)$ with boson-fermion coupling $U_{bf}$ is given as
\begin{align}
    T_2(q,\omega)=\frac{U_{bf}}{1-U_{bf}\Pi_{bf}(q,\omega)},
\end{align}
where we neglect the medium correction to the propagator at the low-energy and weak-interaction limits for approximation, and then consider the in-vacuum two-body propagator as
\begin{align}
    \Pi_{bf}(q,\omega)=
    \sum_{p}\frac{1}{\omega_+-\frac{(p+q/2)^2}{2m_b}-\frac{(-p+q/2)^2}{2m_f}}.
\end{align}
We note that $T_2(q,\omega)$ actually does not explicitly depend on incoming and outgoing relative momenta since we adopt the contact-type interaction here.
Consequently, we obtain $U_{bf}=T_{2}(0,0)=-{1}/(m_ra_{bf})$.}
Besides that, in Eqs.~(1) and (2), $v_{\ell}=2  \ell \pi T $ is the bosonic Matsubara frequency, $a_{bf}$ is the $s$-wave boson-fermion scattering length, $m_r={m_bm_f}/({m_f+m_b})$ is the reduced mass, and $G\left({p}, i \omega_n\right)=\left(i \omega_n-\xi_p \right)^{-1}$ is the thermal Green's function of a fermion with the
fermionic Matsubara frequency $\omega_n=(2n+1)\pi T$ and the {fermionic} kinetic energy {$\xi_{f,p}={p^2}/{(2m_f)}-E_{\rm F}$} measured from the Fermi energy $E_{\rm F}$; $\left({q}, i v_{\ell}\right)=\left({k}-{k}^{\prime}, i v_s-i v_{s^{\prime}}\right)$, ${k}={k}_1-{k}_2, {q}={q}_1-{q}_2, i v_{\ell}=i v_{s_1}-i v_{s_2}$, and $i v_u=$ $i v_{j_1}-i v_{j_2}$ are the transferred momenta.

At the low-energy ($i v_{\ell}=i v_u=0$) and long-wavelength (${q}\rightarrow0$) limits, the medium-induced two- and three-body interaction strengths are obtained as
\begin{align}
    V_2=\,&
    -\frac{m_f}{2\pi k_{\rm F}}\left(\frac{1}{m_ra_{bf}}\right)^2
    ,\\
    V_3=\,&
    \frac{m_f^2}{2\pi k_{\rm F}^3}\left(\frac{1}{m_ra_{bf}}\right)^3,
\end{align}
in one dimension, respectively,
where $k_{\rm F}=\sqrt{2m_fE_{\rm F}}$ is the Fermi momentum.
It is known that the three-body force becomes important in one-dimensional systems~\cite{Drut2018PhysRevLett.120.243002,PhysRevA.97.061605,Sekino2021Phys.Rev.A103.043307,PhysRevA.102.053304}.
{
We note that the medium-induced interactions obtained here decrease as $a_{bf}$ increases, which just corresponds to the peculiarity of the present specific dimensionality associated with the form of the regularized interaction, $T_2(0,0)=-1/(m_r a_{bf})$~\cite{MISTAKIDIS20231}.
}

In addition, such medium-induced interactions generally have momentum dependence with a typical momentum scale $k_{\rm F}$. 
Here, for simplicity, we consider the sufficiently large $k_{\rm F}$ such that the medium-induced interactions can be assumed to be contact-type ones at the long-wavelength and low-energy limits.
Also, we ignore the effective mass and the quasiparticle residue of bosons immersed in the Fermi sea.
{
These may affect the properties of multipolaron bound states; however, in the weak-interaction regime to be considered in this work [i.e., $1/(k_{\rm F}a_{bf})<1$], it is expected to be insignificant.
As a typical instance, the effective mass $m^\ast$ is almost equal to the bare mass $m$ when $1/(k_{\rm F}a_{bf})<1$, as shown in Refs.~\cite{McGuire1966J.Math.Phys.7.123--132,Mao2016Phys.Rev.A94.043645}
, which indicates that the qualitative behavior would be unchanged~\cite{tajima2023polaronic}.
Anyway, the enhanced effective mass helps the cluster formation~\cite{Moriya2021Phys.Rev.C104.065801} and thus the present scheme is sufficient to predict medium-induced clusters.
}

Consequently, the effective Hamiltonian of the system reads
\begin{align}
H=H_0+U_2+U_{\rm{eff}}^{(2)}+U_{\rm{eff}}^{(3)}. 
\end{align}
where $H_0$, $U_2$, and $U_{\rm{eff}}^{(2,3)}$
are the kinetic term of bosons, the boson-boson interaction, and fermion-mediated two- and three-body interactions, respectively.
Each term is given by
\begin{align}
H_0=\,&{\sum_{k} \xi_{b,k} b_k^\dagger b_k}, \\
U_2
=\,&\frac{U_0}{2} \sum_{p_1, p_2, p'_1, p'_2}
b_{p_1}^{\dagger} b_{p_2}^{\dagger} 
b_{p'_2} b_{p'_1} 
\delta_{p_1+p_2, p'_1+p'_2}, \\
U_{\rm{eff}}^{(2)}=\,&
\frac{V_2}{2} \sum_{k, k', q} 
b_{k+q}^{\dagger} b_{k'-q}^{\dagger} 
b_{k'} b_{k}, \\
U_{\rm{eff}}^{(3)}=\,&
\frac{V_3}{6} \sum_{p_1, p_2, p_3, k, q}  
b_{p_1-k-q / 2}^{\dagger} b_{p_2+k-q / 2}^{\dagger} b_{p_3+q}^{\dagger} 
\nonumber\\
&\times b_{p_3} b_{p_2} b_{p_1},
\end{align}
where {bosonic kinetic energy reads $\xi_{b,p}={p^2}/{(2m_b)}$ and} $b_k^{\dagger}$ ($b_k$) is the creation (annihilation) operator of a boson. 
{The bosonic contact coupling $U_0$ can be expressed in terms of the $s$-wave boson-boson scattering length $a_{bb}$ as
\begin{align}
    U_0=-\frac{2}{m_ba_{bb}},
\end{align}
which can be derived from the two-body $T$ matrix as
\begin{align}
    \frac{1}{U_0}-\sum_{p}\frac{m_b}{k^2+i\delta-p^2}
    =\frac{m_b}{2}\left(-{a_{bb}}+\frac{i}{k}\right),
\end{align}
where $\delta$ is an infinitesimally small number.
Here the bosonic interaction $U_0$ is an attractive (repulsive) one when the scattering length $a_{bb}$ is positive (negative).
In addition, we note that the ultraviolet divergence is absent in the present one-dimensional case, which is different from the three-dimensional one~\cite{Tajima2021Atoms9.18}.}

{As for the application to the quantum simulation for the alpha cluster, the present medium fermions and impurity bosons correspond to the extra neutrons and alpha clusters, respectively, which is similar to the alpha cluster model~\cite{Phyu2020Prog.Theor.Exp.Phys.2020.093D01,Moriya2021Phys.Rev.C104.065801}.
However, for the specific Hamiltonian of the alpha clustering, since many other contributions such as the detail of alpha-alpha interaction and Coulomb repulsion are needed to be taken into consideration, it will be much more complicated than the present model setup and left for a future work.
Consequently, here, as a first-step study, we consider the simplified bosonic interaction characterized by the $s$-wave scattering length for the sake of the elucidation of effects of fermion-mediated interaction on cluster formations.}

To see three-body bosonic bound states,
we consider the zero center-of-mass frame of three bosons.
The three-body variational function is adopted as
\begin{align}
|\Psi\rangle=\,&
\sum_{k_1, k_2, k_3} 
\Omega_{k_1, k_2} 
b_{k_1}^{\dagger} 
b_{k_2}^{\dagger} 
b_{k_3}^{\dagger} 
\delta_{k_1+k_2+k_3,0}|0\rangle\nonumber\\
=\,&
\sum_{k_1, k_2} 
\Omega_{k_1, k_2} 
A_{{k_1, k_2}}^{\dagger} 
|0\rangle,
\end{align}
where the trimer creation and annihilation operators at zero center-of-mass momentum are defined as
\begin{align}
\label{eq:a}
A_{{k_1, k_2}}^{\dagger} 
&=
\sum_{k_3}
b_{k_1}^{\dagger} 
b_{k_2}^{\dagger} 
b_{k_3}^{\dagger}
\delta_{k_1+k_2+k_3, 0}
,\nonumber\\ 
A_{{k_1, k_2}} 
&=
\sum_{k_3}
b_{k_3}
b_{k_2}
b_{k_1}
\delta_{k_1+k_2+k_3, 0},
\end{align}
{and the reference vacuum state $|0\rangle$ is defined as {the ground state of the fermions without bosons.}}
Because of the commutation relation of bosonic operators in Eq.~\eqref{eq:a}, one can find
\begin{align}
\Omega_{k_1, k_2} =\Omega_{k_2, k_1} .  
\end{align}

From the variational principle, we obtain
\begin{align}
\frac{\delta\langle\Psi_3|(H-E_3)|\Psi_3\rangle}{\delta\Omega_{p_1,p_2}^*}=0,
\end{align}
where $E_3$ is the three-body ground-state energy.
The resulting variational equation reads
\begin{widetext}
\begin{align}\label{eqv3}
&
(\xi_{k_1}+\xi_{k_2}+\xi_{-k_1-k_2}-E_3)
(\Omega_{k_1, k_2}+\Omega_{k_2,-k_1-k_2}+\Omega_{-k_1-k_2,k_1})\nonumber\\
=\,&
-(U_0+V_2) 
\sum_{q} 
\left(
2\Omega_{-k_1-k_2,q}
+
\Omega_{q, k_1+k_2-q}
+
2\Omega_{k_2, q}
+
\Omega_{q, -q-k_2}
+
2\Omega_{k_1, q}
+
\Omega_{q, -q-k_1}
\right)
-
3V_3
\sum_{p_1,p_2} 
\Omega_{p_1, p_2}.
\end{align}
Detailed expressions for the expectation values of each term in the Hamiltonian can be found in the Appendix~\ref{appendixA}.

\section{Results and discussion}\label{sec:III}

\subsection{In-vacuum case with $V_2=V_3=0$}\label{sec:IIIA}

For the first step, we consider the in-vacuum case with vanishing fermion-mediated two- and three-body interactions.
As a result, the variational equation is simplified as
\begin{align}\label{eq20}
&
(\xi_{k_1}+\xi_{k_2}+\xi_{-k_1-k_2}-E_3)
(\Omega_{k_1, k_2}+\Omega_{k_2,-k_1-k_2}+\Omega_{-k_1-k_2,k_1})\nonumber\\
=\,&
-U_0 
\sum_{q} 
\left(
2\Omega_{-k_1-k_2,q}
+
\Omega_{q, k_1+k_2-q}
+
2\Omega_{k_2, q}
+
\Omega_{q, -q-k_2}
+
2\Omega_{k_1, q}
+
\Omega_{q, -q-k_1}
\right).
\end{align}
\end{widetext}
Here we further introduce 
\begin{align}
    \mathcal{B}(p)=\sum_q
    (
    2\Omega_{p,q}
+
\Omega_{q, -p-q}
    ).
\end{align}
Using $\mathcal{B}(p)$, we rewrite Eq.~\eqref{eq20} as
\begin{align}
&(\xi_{k_1}+\xi_{k_2}+\xi_{-k_1-k_2}-E_3)\nonumber\\
&\times(\Omega_{k_1, k_2}+\Omega_{k_2,-k_1-k_2}+\Omega_{-k_1-k_2,k_1})\nonumber\\
=\,&
-U_0 
[\mathcal{B}(-k_1-k_2)+\mathcal{B}(k_2)+\mathcal{B}(k_1)].
\end{align}
The set of the variational parameters $\Omega_{k_1, k_2}+\Omega_{k_2,-k_1-k_2}+\Omega_{-k_1-k_2,k_1}$ can be expressed as
\begin{align}
&\Omega_{k_1, k_2}+\Omega_{k_2,-k_1-k_2}+\Omega_{-k_1-k_2,k_1}\nonumber\\
=\,&
-U_0 
\frac{
\mathcal{B}(-k_1-k_2)+\mathcal{B}(k_2)+\mathcal{B}(k_1)}
{
\xi_{k_1}+\xi_{k_2}+\xi_{-k_1-k_2}-E_3}.
\end{align}
Consequently, the self-consistent equation for $\mathcal{B}(p)$ reads
\begin{align}\label{eq3bv0}
&\mathcal{B}(k_2)
\left(
1+U_0 \sum_{k_1}
\frac{1}
{
\xi_{k_1}+\xi_{k_2}+\xi_{-k_1-k_2}-E_3}
\right)\nonumber\\
=\,&
-2U_0 \sum_{k_1}
\frac{
\mathcal{B}(k_1)}
{
\xi_{k_1}+\xi_{k_2}+\xi_{-k_1-k_2}-E_3}.
\end{align}
The momentum summation in the left-hand side of Eq.~\eqref{eq3bv0} corresponds to the contribution associated with the two-body scattering process.
It can be analytically evaluated as
\begin{align}\label{inf}
   & U_0 \sum_{k_1}
\frac{1}
{
\xi_{k_1}+\xi_{k_2}+\xi_{-k_1-k_2}-E_3}\nonumber\\
=\,&\frac{mU_0}{\pi}
\left(
\frac{\textrm{arctan}\frac{2\Lambda_{{2}}+k_2}{\sqrt{3k_2^2-4mE_3}}}{\sqrt{3k_2^2-4mE_3}}
-
\frac{\textrm{arctan}\frac{-2\Lambda_{{2}}+k_2}{\sqrt{3k_2^2-4mE_3}}}{\sqrt{3k_2^2-4mE_3}}\right)
\nonumber\\
=\,&\frac{mU_0}{\sqrt{3k_2^2-4mE_3}}
\quad(\Lambda_{{2}}\rightarrow\infty).
\end{align}
In the last expression of Eq.~\eqref{inf}, we take an infinitely large cutoff $\Lambda_2\rightarrow\infty$ for the two-body scattering process.

\begin{figure}
  \includegraphics[width=0.45\textwidth]{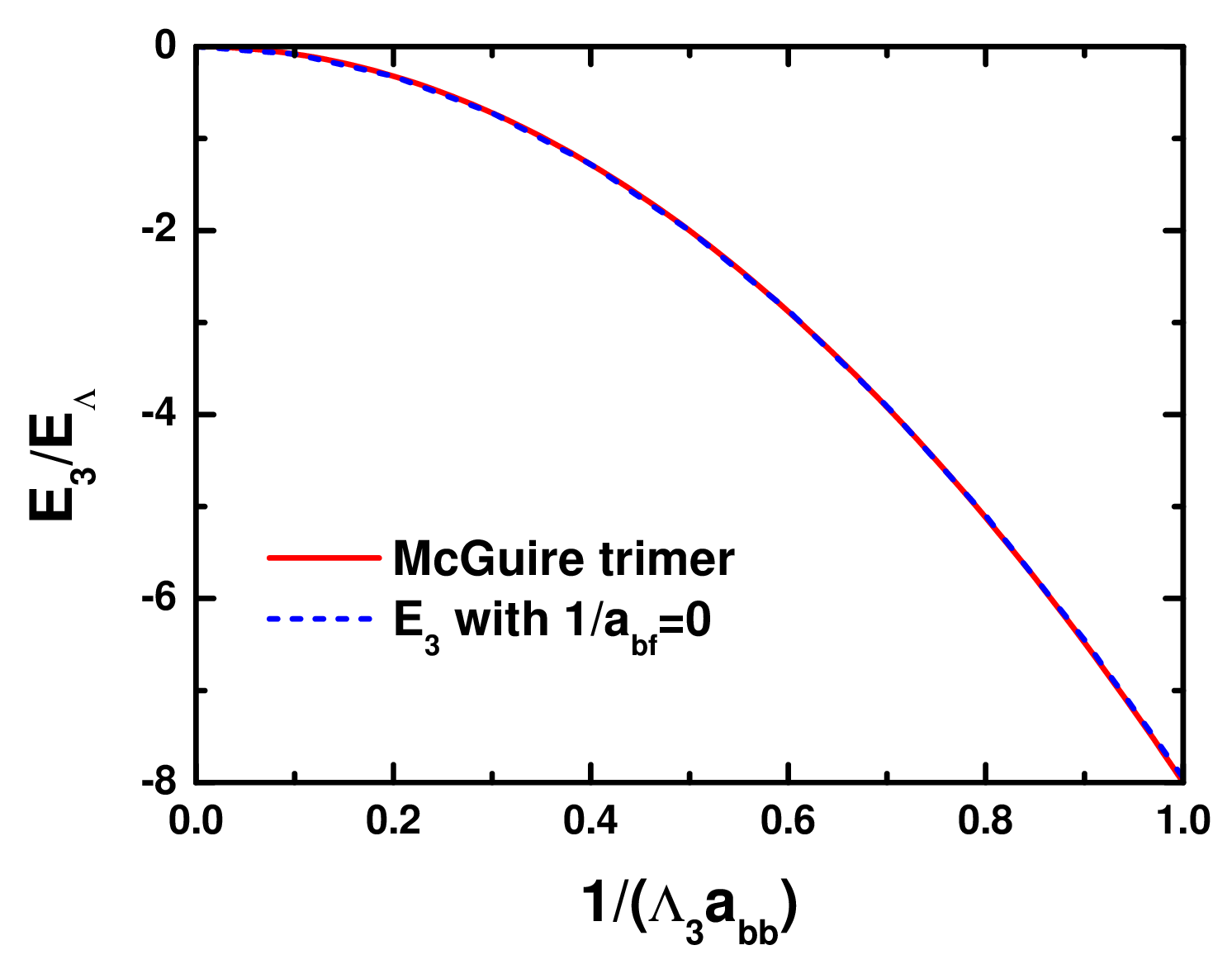}
  \caption{{Three-body ground-state energy $E_3$ as a function of the inverse scattering length $1/(\Lambda_3 a_{bb})$ with the momentum cutoff $\Lambda_3$ and energy scale $E_{\Lambda}=\Lambda_3^2/(2m_b)$.
 {For comparison, the exact result of the McGuire trimer energy is also plotted~\cite{McGuire2004J.Math.Phys.5.622--636}.}
  }
  }\label{fig:b1}
\end{figure}

{We obtain the solution of three-body ground states by solving Eq.~\eqref{eq3bv0}.
The numerical result of the three-body energy $E_3$ is shown in Fig.~\ref{fig:b1} as a function of the inverse bosonic scattering length $1/a_{bb}$, where the momentum cutoff $\Lambda_3$ for the three-body sector is employed.
In order to achieve the infinite cutoff limit, a sufficiently large number of steps for the discretization is performed.
The McGuire result of an exact trimer energy~\cite{McGuire2004J.Math.Phys.5.622--636}, where its energy is given by $E_3=-4/(m_b a_{bb}^2)$, is also plotted for comparison.
It can be seen that the solution of the three-body ground state in vacuum agrees well with the McGuire trimer result.
These three-body bound states are stable compared to the two-body ones with the binding energy $1/(m_ba_{bb}^2)$, although we do not explicitly show it here.}

\subsection{Bosonic clusters arising from medium-induced two- and three-body interactions}\label{sec:IIIB}

In this subsection, let us investigate the bosonic clusters with medium-induced two- and three-body interactions.
From the full variational equation~\eqref{eqv3}, one has
\begin{align}\label{eq27}
&
\Omega_{k_1, k_2}+\Omega_{k_2,-k_1-k_2}+\Omega_{-k_1-k_2,k_1}\nonumber\\
=\,&
-(U_0+V_2) 
\frac{
 \mathcal{B}(-k_1-k_2)
+
 \mathcal{B}(k_2)
+
 \mathcal{B}(k_1)
}{\xi_{k_1}+\xi_{k_2}+\xi_{-k_1-k_2}-E_3}\nonumber\\
&-
3V_3
\frac{\mathcal{C}}{\xi_{k_1}+\xi_{k_2}+\xi_{-k_1-k_2}-E_3},
\end{align}
where we introduced
\begin{align}
\label{eq:c}
    \mathcal{C}=\sum_{p_1,p_2}\Omega_{p_1,p_2}.
\end{align}
Further taking the summation of $k_1$ in Eq.~\eqref{eq20}, one obtains
\begin{align}\label{eq25}
&\mathcal{B}(k_2)
\left[1+(U_0+V_2)  \sum_{k_1}
\frac{
1
}{\xi_{k_1}+\xi_{k_2}+\xi_{-k_1-k_2}-E_3}\right]\nonumber\\
=\,&
-(U_0+V_2)  \sum_{k_1}
\frac{
2\mathcal{B}(k_1)
}{\xi_{k_1}+\xi_{k_2}+\xi_{-k_1-k_2}-E_3}\nonumber\\
&-
3V_3\sum_{k_1}
\frac{
\mathcal{C}
}{\xi_{k_1}+\xi_{k_2}+\xi_{-k_1-k_2}-E_3}.
\end{align}
{We note that here Eq.~\eqref{eq25} actually reproduces {the Skorniakov-Ter-Martirosian equation~\cite{Naidon2017Rep.Prog.Phys.80.056001} in the case with the absence of the medium-induced interactions.}} 
On the other hand, taking the summation of $k_1$ and $k_2$ in Eq.~\eqref{eq27}, one has
\begin{align}
\mathcal{C}
&=
- \left[1+
V_3\sum_{k_1,k_2}
\frac{1}{\xi_{k_1}+\xi_{k_2}+\xi_{-k_1-k_2}-E_3}\right]^{-1}\cr
&\times
\frac{U_0+V_2}{3}
\sum_{k_1,k_2}
\frac{
 \mathcal{B}(-k_1-k_2)
+
 \mathcal{B}(k_2)
+
 \mathcal{B}(k_1)
}{\xi_{k_1}+\xi_{k_2}+\xi_{-k_1-k_2}-E_3}.
\end{align}

\begin{figure}[t]
  \includegraphics[width=0.5\textwidth]{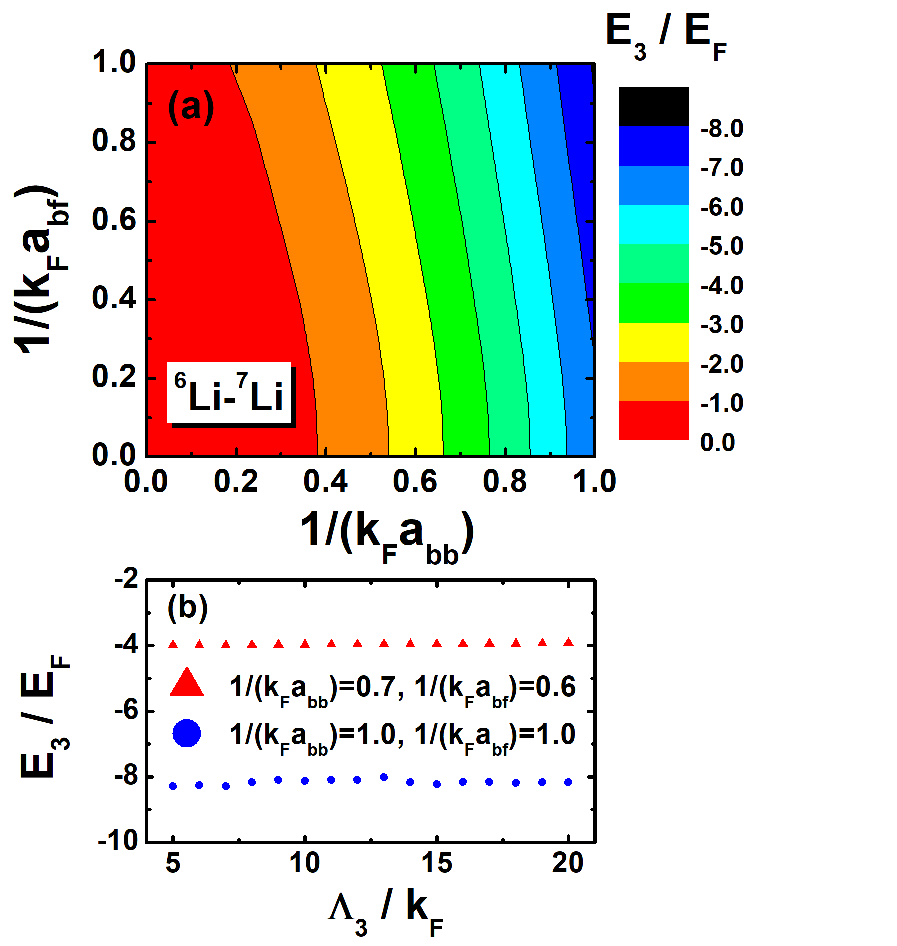}
  \caption{
  (a) Contour plot of in-medium three-body energy $E_3$ in a plane of $1/(k_{\rm F} a_{bb})$ and $1/(k_{\rm F} a_{bf})$ for $^6$Li-$^7$Li mixture.
  The momentum cutoff $\Lambda_3$ for the three-body sector is taken as $10k_{\rm F}$.
  (b) Momentum cutoff dependence of $E_3$ at $1/(k_{\rm F} a_{bb})=0.7$, $1/(k_{\rm F} a_{bf})=0.6$ and $1/(k_{\rm F} a_{bb})=1.0$, $1/(k_{\rm F} a_{bf})=1.0$.
  }\label{fig:b2}
\end{figure}

Consequently, the closed equation for $\mathcal{B}(p)$ and $E_3$ reads
\begin{align}
\label{eq102}
    &\mathcal{B}(k_2)\left[\frac{1}{U_0+V_2}+I_2(k_2,E_3)\right]\nonumber\\
    &\  =I_3(k_2,E_3{; \mathcal{B}})+\frac{I_4(E_3{; \mathcal{B}})I_5(k_2,E_3)}{1+I_6(E_3)},
\end{align}
where we introduced
\begin{align}
    I_2(k_2,E_3)
    =\sum_{k_1}\frac{1}{\xi_{k_1}+\xi_{k_2}+\xi_{-k_1-k_2}-E_3},
\end{align}
\begin{align}
\label{eq:I3}
    I_3(k_2,E_3{; \mathcal{B}})
    =\sum_{k_1}\frac{-2\mathcal{B}(k_1)}{\xi_{k_1}+\xi_{k_2}+\xi_{-k_1-k_2}-E_3},
\end{align}
\begin{align}
\label{eq:I4}
    I_4(E_3{; \mathcal{B}})=\sum_{k_1,k_2}\frac{\mathcal{B}(k_1)}{\xi_{k_1}+\xi_{k_2}+\xi_{-k_1-k_2}-E_3},
\end{align}
\begin{align}
\label{eq:I5}
    I_5(k_2,E_3)={3V_3}\sum_{k_1}\frac{
1}{\xi_{k_1}+\xi_{k_2}+\xi_{-k_1-k_2}-E_3},
\end{align}
and
\begin{align}
\label{eq:I6}
    I_6(E_3)=V_3\sum_{k_1,k_2}
    \frac{1}{\xi_{k_1}+\xi_{k_2}+\xi_{-k_1-k_2}-E_3},
\end{align}
respectively.
 By taking the infinite cutoff limit as done in Eq.~\eqref{inf},
one can analytically perform the momentum summation in $I_2(k_2,E_3)$ and $I_5(k_2,E_3)$ as
\begin{align}
    {I_2(k_2,E_3)}
    =
    \frac{m_b}{{\sqrt{3{k}_2^2-4m_b{E}_3}}},
\end{align}
and
\begin{align}
    I_5(k_2,E_3)=3{V_3}\frac{m_b}{{\sqrt{3{k}_2^2-4m_b{E}_3}}}.
\end{align}

In Fig.~\ref{fig:b2}(a), by considering the $^6$Li-$^7$Li mixture as an example, the in-medium three-body energy $E_3$ obtained from Eq.~\eqref{eq102} is shown in a plane of the inverse scattering lengths $1/(k_{\rm F} a_{bb})$ and $1/(k_{\rm F} a_{bf})$.
 $\Lambda_3$ is taken as $10k_{\rm F}$ for the integrals of the three-body sector given by Eqs.~\eqref{eq:I3}, \eqref{eq:I4}, and \eqref{eq:I6}.
 It can be found that $|E_3|$ becomes larger when $1/(k_{\rm F}a_{ bf})$ is larger, indicating the formation of the medium-assisted three-body bound states.
$\Lambda_3$ dependence of $E_3$ is also shown in Fig.~\ref{fig:b2}(b).
{
With an infinitely large cutoff being taken for the two-body scattering process,}
one can see that our results of $E_3$ are insensitive with respect to the change of $\Lambda_3$.

\begin{figure}[t]
  \includegraphics[width=0.45\textwidth]{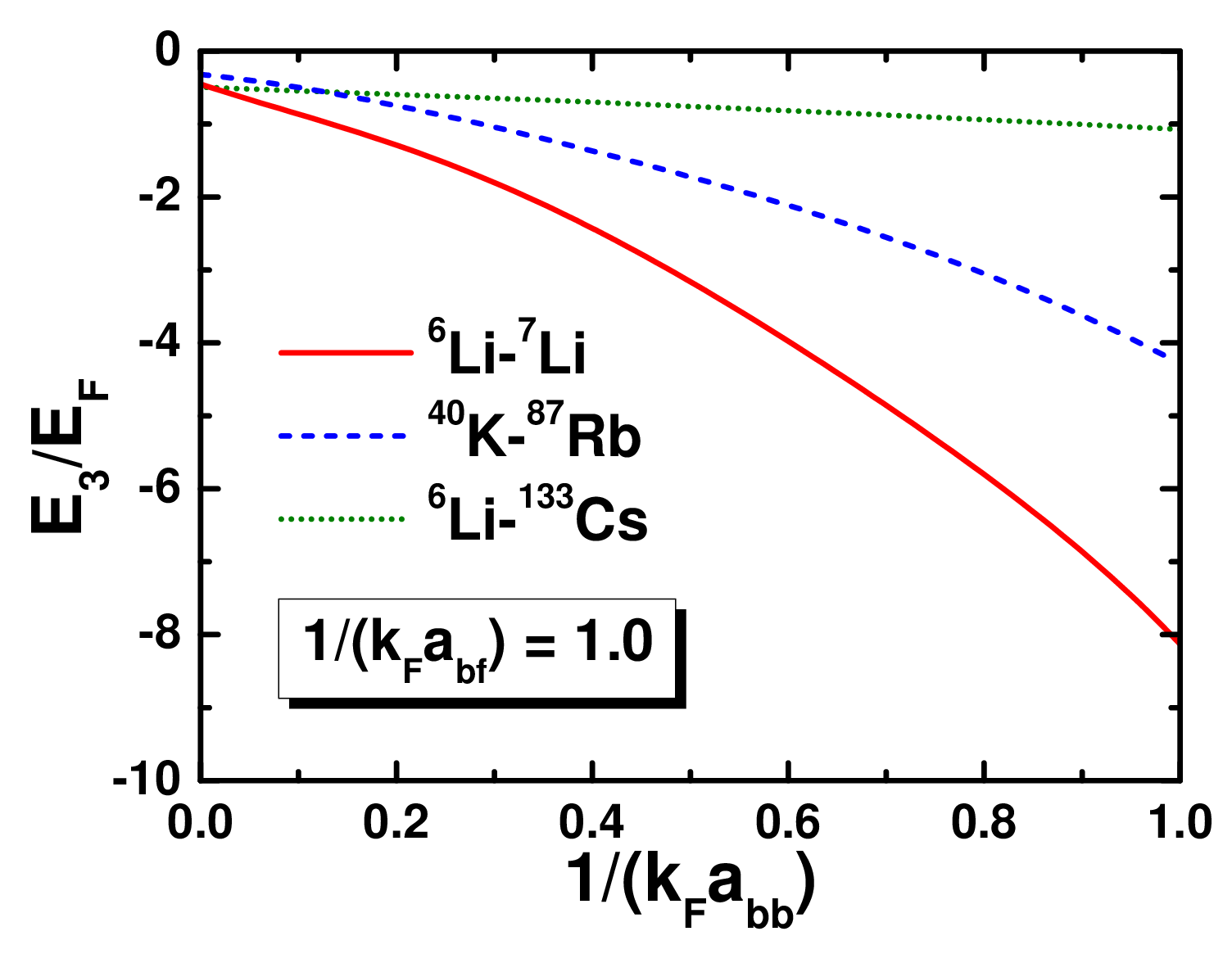}
  \caption{In-medium three-body energy $E_3/E_{\rm F}$ as a function of $1/(k_{\rm F} a_{bb})$ at $1/(k_{\rm F} a_{bf})=1.0$ with different mixtures.
  $\Lambda_3$ is taken as $10k_{\rm F}$.
  }\label{fig:b3}
\end{figure}

While we have discussed the $^6$Li-$^7$Li mixture above, the formation of the in-medium bound states may be affected by the mass difference between the Fermi and Bose atoms.
In Fig.~\ref{fig:b3}, the in-medium three-body energies $E_3$ are shown as functions of $1/(k_{\rm F} a_{bb})$ for $^6$Li-$^7$Li, $^{40}$K-$^{87}$Rb, and $^{6}$Li-$^{133}$Cs mixtures, respectively, with a fixed $s$-wave boson-fermion scattering length, that is, $1/(k_{\rm F} a_{bf})=1.0$.
$\Lambda_3$ is also taken as $10k_{\rm F}$ in Fig.~\ref{fig:b3}.
$|E_3|$ becomes larger with the increase of the inverse bosonic scattering length $1/(k_{\rm F}a_{bb})$, indicating that the in-medium three-body state becomes stabler.
For the mixture with a larger mass ratio $m_b/m_f$ (e.g., $^6$Li-$^{133}$Cs mixture),
$|E_3|$ becomes nonzero even at $1/(k_{\rm F} a_{bb})=0$, but $E_3/E_{\rm F}$ keeps a small value under the increase of $1/(k_{\rm F}a_{bb})$ compared to the nearly mass-balanced case (i.e., $^6$Li-$^7$Li mixture).
We note that this is due to the modification of the fermionic energy scale in the dimensionless ratio, 
\begin{align}
\label{eq:dim}
    E_3/E_{\rm F}\equiv\frac{\kappa^2}{2m_b}\frac{2m_f}{k_{\rm F}^2}=\frac{\kappa^2}{k_{\rm F}^2}\frac{m_f}{m_b},
\end{align}
where we introduced the effective three-body parameter $\kappa$~\cite{Naidon2017Rep.Prog.Phys.80.056001} {as a momentum scale for $E_3$, although the Efimov state, for which $\kappa$ is conventionally used in three dimensions (3D), is absent in the present 1D system.}

\begin{figure}
  \includegraphics[width=0.45\textwidth]{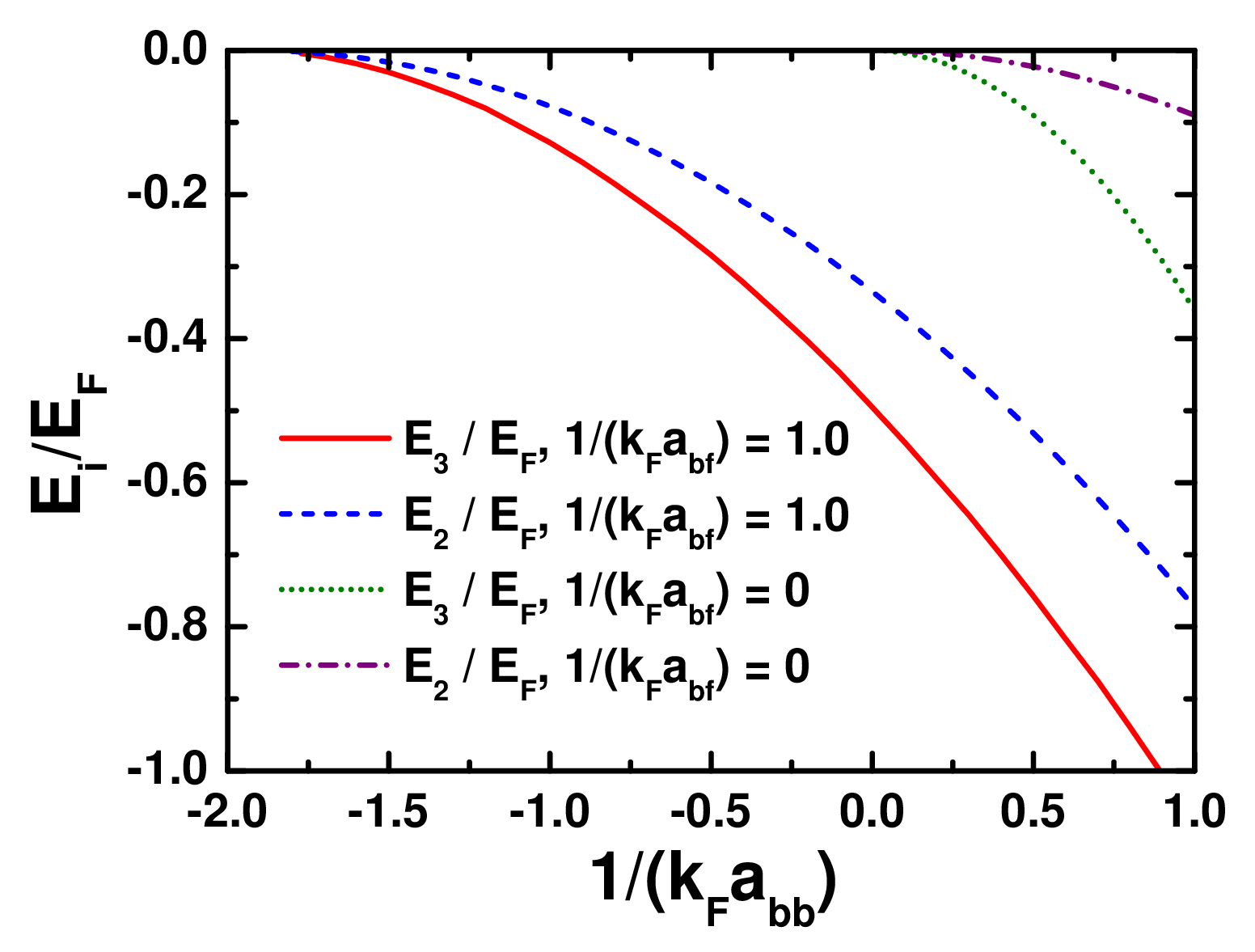}
  \caption{In-medium two- and three-body energies of the $^6$Li-$^{133}$Cs mixture as functions of $1/(k_{\rm F} a_{bb})$ at $1/(k_{\rm F} a_{bf})=1.0$ for the whole regime.
$\Lambda_3$ is taken as $10k_{\rm F}$.
  Two- and three-body energies with $1/(k_{\rm F} a_{bf})=0$ are also shown for comparison.
  }\label{fig:abb}
\end{figure}

As a step further, the in-medium three-body energy $E_3/E_{\rm F}$ in a $^6$Li-$^{133}$Cs mixture is plotted in Fig.~\ref{fig:abb} as a function of $1/(k_{\rm F} a_{bb})$ with fixed $1/(k_{\rm F}a_{bf})$.
It is found that $E_3$ is nonzero even for negative $a_{bb}$ and eventually becomes zero at a certain negative value of $a_{bb}$.
Namely, the bosonic three-body clusters can be formed even with a repulsive boson-boson interaction due to the fermion-mediated interactions.
This result is in sharp contrast with the case with $1/(k_{\rm F}a_{bf})=0$, which exactly reproduces the McGuire trimer results and becomes nonzero only when $a_{bb}>0$.
For comparison, here we also show the two-body energy with fermion-mediated interactions.
For the case with $1/a_{bf} =0$, the two-body energy $E_{2,0}$ can be obtained as
    $E_{2,0}= -\frac{1}{m_b a_{bb}^2}$.
In the presence of finite $a_{bf}$, {namely, by replacing $U_0$ with $U_0+V_2$,} here we introduce the effective boson-boson scattering length as
\begin{align}
     \frac{1}{{k_{\rm F}}a_{bb}^{\rm eff}}
     =
     \frac{1}{{k_{\rm F}}a_{bb}}
    +\frac{m_fm_b}{4\pi m_r^2}\left(\frac{1}{k_{\rm F}a_{bf}}\right)^2.
\end{align}
Consequently, one has the in-medium two-body energy as
\begin{align}\label{e22}
    E_2 =  -\frac{1}{m_b {a_{bb}^{\rm eff}}^2}.
\end{align}
We show such a two-body ground state as the blue dashed line in Fig.~\ref{fig:abb}.
$E_2$ also becomes nonzero simultaneously with $E_3$ at $1/(k_{\rm F}a_{bb})\simeq -1.8$ in the case with $1/(k_{\rm F}a_{bf})=1.0$.
Indeed, the same tendency can be found in the case with $1/(k_{\rm F}a_{bf})=0$.
In this regard, our results indicate that the formation of the in-medium bosonic clusters can be observed via the medium shift of three-body resonances (corresponding to the threshold for the three-body recombination).
\begin{figure}[t]
  \includegraphics[width=0.45\textwidth]{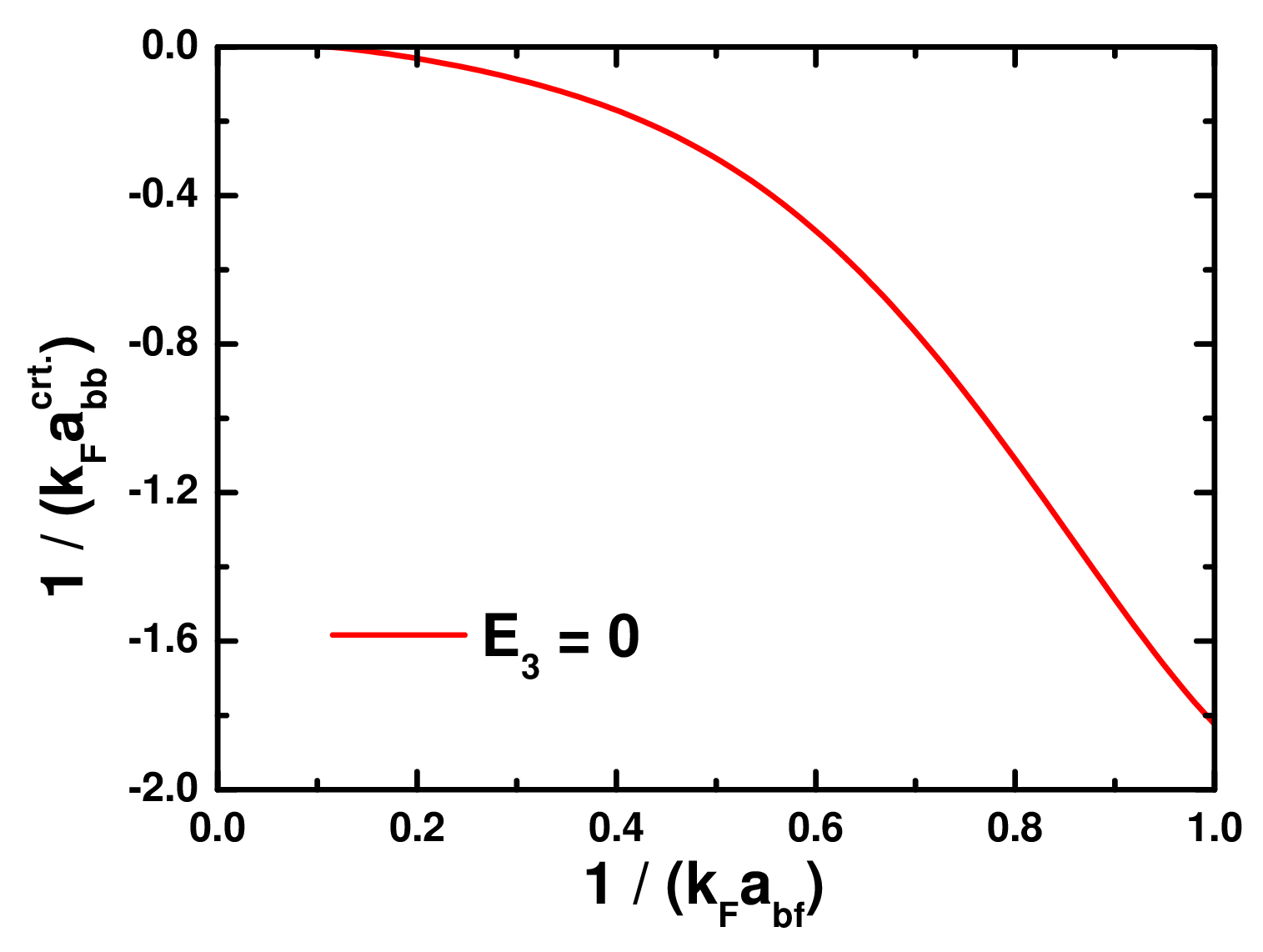}
  \caption{
  {
  In-medium critical bosonic coupling $1/(k_{\rm F}a^{\rm crt.}_{bb})$, where one has $E_3=0$, as a function of $1/(k_{\rm F}a_{bf})$.
  }
  }\label{fig:abf}
\end{figure}
{Based on Ref.~\cite{Braaten2017Phys.Rev.A95.012708}, the three-body loss rate is related to $\mathcal{C}$ in Eq.~\eqref{eq:c}, which characterizes the amplitude of the three-body wave function.
In the present work, we numerically check that $\mathcal{C}$ becomes larger estimated from the difference of the couplings of $E_3=0$ with and without medium (i.e., Fermi sea).
For a $^6$Li-$^{133}$Cs mixture, we also show the in-medium critical bosonic interaction strength $1/(k_{\rm F}a^{\rm crt.}_{bb})$, where $E_3=0$, as a function of $1/(k_{\rm F}a_{bf})$ in Fig.~\ref{fig:abf}.
It can be seen that with stronger boson-fermion interaction strength $1/(k_{\rm F}a_{bf})$, the shift of critical bosonic coupling also increases.
Note that $1/(k_{\rm F}a_{bb}^{\rm crt.})=0$ is found in the absence of the Fermi sea where the McGuire trimer appears, as shown in Fig.~\ref{fig:b1}.
The three-body loss rate can thus be connected to the medium shift of the three-body loss threshold.
Incidentally, the spectral function of Bose polarons under the presence of Efimov three-body correlations, which is similar to our setting for impurities in a Fermi sea, has been calculated in the high-temperature regime via the virial expansion~\cite{Sun2017Phys.Rev.Lett.119.013401}.
As visible signatures of Efimov branches can be found in such a similar system, additional branches associated with in-medium clusters can also be expected to be found in the spectral function. 
However, the calculation of the spectral function in the present case is left for future work.}

\begin{figure}[t]
  \includegraphics[width=0.45\textwidth]{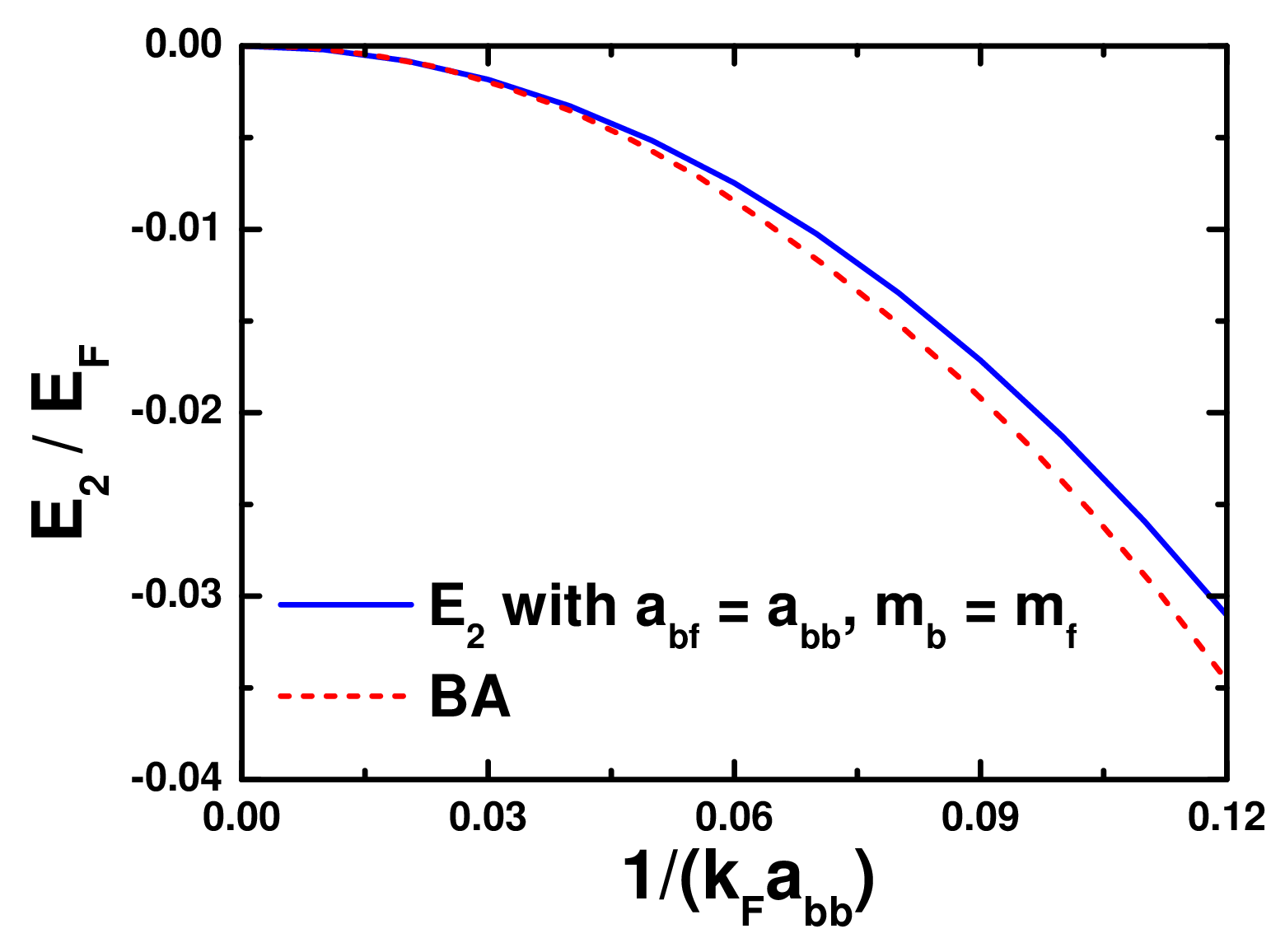}
  \caption{
  {
  In-medium two-body energy $E_2/E_{\rm F}$ as a function of $1/(k_{\rm F} a_{bb})$ with $1/(k_{\rm F} a_{bf})=1/(k_{\rm F} a_{bb})$ and $m_b=m_f$.
  The one calculated using the Bethe ansatz~\cite{PhysRevResearch.1.033177} is also shown for comparison.
  }
  }\label{fig:ba}
\end{figure}

{Here we also show the in-medium two-body energy {$E_2$} with $1/(k_{\rm F} a_{bf})=1/(k_{\rm F} a_{bb})$ and $m_b=m_f$ in Fig.~\ref{fig:ba}.
It can be seen that the qualitative behavior of {$E_2$} calculated via Eq.~\eqref{e22} is similar to the one obtained using the Bethe ansatz in Ref.~\cite{PhysRevResearch.1.033177}, which also indicates that the qualitative behavior of the multipolaron bound states will not change under our present approximation at low-energy and weak-coupling limits.}

\begin{figure}[t]
  \includegraphics[width=0.45\textwidth]{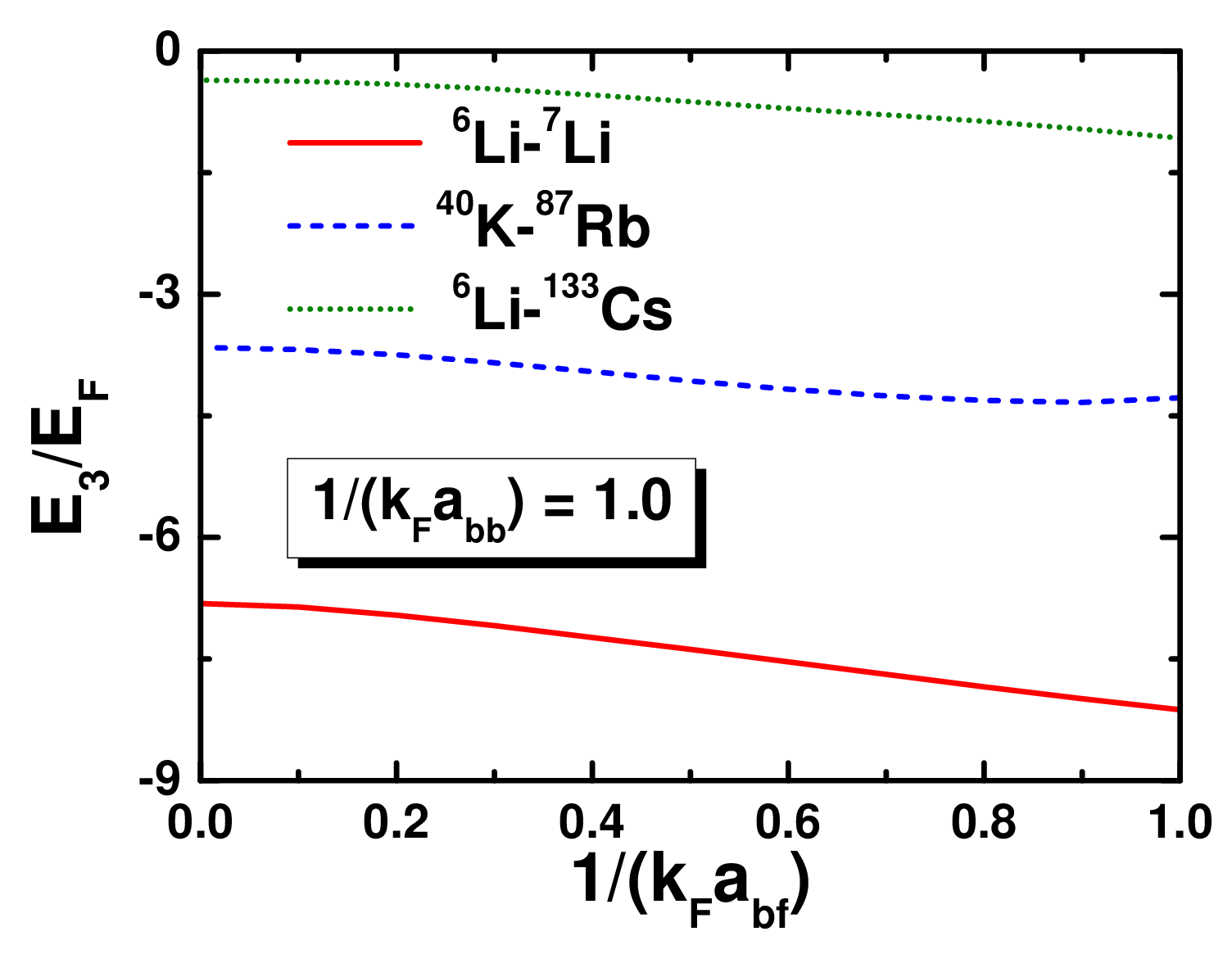}
  \caption{In-medium three-body energies $E_3/E_{\rm F}$ as functions of $1/(k_{\rm F} a_{bf})$ at $1/(k_{\rm F} a_{bb})=1.0$ with different mixtures.
 $\Lambda_3$ is taken as $10k_{\rm F}$.
  }\label{fig:b4}
\end{figure}

Similarly, in Fig.~\ref{fig:b4},
we also plot $E_3/E_{\rm F}$ with fixed $1/(k_{\rm F} a_{bb})=1.0$ as functions of $1/(k_{\rm F} a_{bf})$ in $^6$Li-$^7$Li, $^{40}$K-$^{87}$Rb, and $^{6}$Li-$^{133}$Cs mixtures, respectively.
At vanishing medium-induced interaction, namely, $1/(k_{\rm F} a_{bf})=0$,
$E_3/E_{\rm F}$ is proportional to the ratio $m_f/m_b$ as found in Eq.~\eqref{eq:dim}.
With the increase of boson-fermion coupling strength $1/(k_{\rm F} a_{bf})$, $E_3$ monotonically decreases due to the fermion-mediated interactions.
Such behavior is qualitatively unchanged for each mixture.

\begin{figure}[t]
  \includegraphics[width=0.45\textwidth]{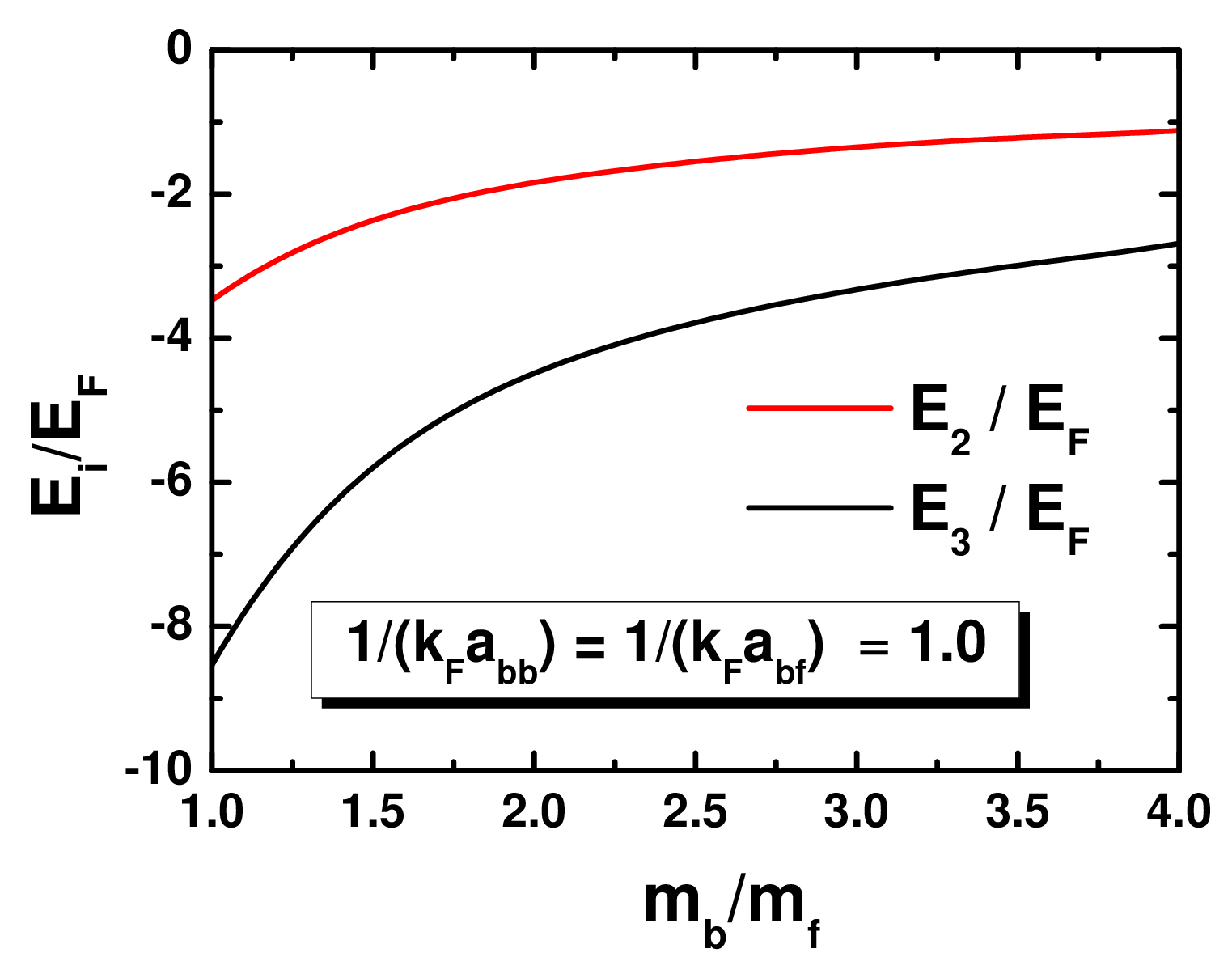}
  \caption{In-medium two- and three-body energies as functions of mass ratio $m_b/m_f$ at $1/(k_{\rm F} a_{bf})=1/(k_{\rm F} a_{bb})=1.0$.
  $\Lambda_3$ is taken as $10k_{\rm F}$.
  }\label{fig:b5}
\end{figure}

Finally, at $1/(k_{\rm F} a_{bf})=1/(k_{\rm F} a_{bb})=1.0$, both the in-medium two- and three-body energies are shown as functions of mass ratio $m_b/m_f$ in Fig.~\ref{fig:b5}.
It can be seen that $|E_3|$ is larger than $|E_2|$ for an arbitrary mass ratio, at least in the region we explored.
In this sense, while we considered specific mixtures such as $^6$Li-$^7$Li, $^{40}$K-$^{87}$Rb, and $^6$Li-$^{133}$Cs mixtures, our result can be useful for other mixtures with a different mass ratio.
In particular, the result at $m_b/m_f=4$ is relevant for alpha clusters in neutron-rich matter because the mass of an alpha particle consisting of four nucleons may be approximately given by the quadruple nucleon masses.
In such a case, $|E_2|$ and $|E_3|$ may be regarded as the in-medium binding energies of $^8$Be and $^{12}$C nuclei in neutron matter~\cite{Moriya2021Phys.Rev.C104.065801}.
The fermionic spin degrees of freedom may further enhance the binding of these clusters because the fermion-mediated interaction can increase.

We note that the increase of $E_{2,3}/E_{\rm F}$ with increasing $m_b/m_f$ is also associated with the change of $E_{\rm F}$, which is proportional to $m_f^{-1}$, as we found in Eq.~\eqref{eq:dim}. 
For the two-body case, one can find
    $\frac{E_2}{E_{\rm F}}=-2\frac{m_f}{m_b}\frac{1}{(k_{\rm F}a_{bb}^{\rm eff})^2}$,
which indicates that $E_2/E_{\rm F}$ can increase with $m_b/m_f$ even for fixed $1/(k_{\rm F}a_{\rm bb}^{\rm eff})$, while noting that $1/(k_{\rm F}a_{\rm bb}^{\rm eff})$ also depends on $m_b/m_f$.

\section{Summary and perspectives}\label{sec:IV}

In this paper, we have investigated the cluster formation of bosonic atoms immersed in a dilute background Fermi gas in one dimension.
To this end, we have employed the variational approach and confirmed that it can reproduce an exact three-body result of the McGuire trimer~\cite{McGuire2004J.Math.Phys.5.622--636} in vacuum.
After that, we have numerically calculated the two- and three-body binding energies of bosonic clusters associated with the medium-induced two- and three-body interactions in this system.
In addition to the stabilization of these clusters due to the fermion-mediated interaction,
it is found that the bosonic clusters can be formed even with a repulsive boson-boson interaction, in contrast to the in-vacuum case.
Our work would indicate that the in-medium cluster formation can be observed via the
three-body recombination process in ultracold-atom experiments.
In addition, our setup is similar to the previous work for $\alpha$ particles in cold neutron matter in three dimensions~\cite{Moriya2021Phys.Rev.C104.065801}. 
Consequently, our study also proposes a way to perform the quantum simulation of $\alpha$ clustering in neutron-rich matter by using an ultracold-atomic system.  

For future perspectives, while we employed the one-dimensional system for simplicity, it would be important to consider higher-dimensional systems.
Moreover, we assumed the contact-type medium-induced interactions, but their nonlocal properties may be important for more quantitative studies.
The effect of the polaronic effective mass~\cite{Moriya2021Phys.Rev.C104.065801,tajima2023polaronic} may also be worth investigating.  
The larger clustering states in medium such as tetramer and pentamer are also left for an interesting future work.

\begin{acknowledgments}
The authors are grateful to Artem Volosniev for providing the numerical data calculated using the Bethe ansatz as shown in Fig.~\ref{fig:ba}.
Y.~G. was supported by the RIKEN Junior Research Associate Program.
H.~T. thanks Pascal Naidon, Munekazu Horikoshi, Takahiro M. Doi, Hajime Moriya, Wataru Horiuchi, Eiji Nakano, and Kei Iida 
for the useful discussion.
H.~T. acknowledges the JSPS Grants-in-Aid for Scientific Research under Grants No.~18H05406, No.~22H01158, and No.~22K13981.
\end{acknowledgments}

\appendix

\begin{widetext}
\section{Expectation value of Hamiltonian}\label{appendixA}

In this appendix, we show the detailed expressions for the expectation values of each term in the Hamiltonian.
By applying the trial wave function on the Hamiltonian, the expectation values for each part reads
\begin{align}
\left\langle H_0\right\rangle
=\,&
\sum_{k} 
\sum_{k_1, k_2} 
\sum_{k'_1, k'_2} 
\Omega_{k_1, k_2}^* 
\Omega_{k'_1, k'_2}
\left\langle 0\right|
A_{k_1, k_2}
\xi_k
b_k^{\dagger} 
b_k
A_{k'_1, k'_2}^{\dagger} 
\left|0\right\rangle 
\delta_{k_1+k_2+k_3, 0}
\delta_{k'_1+k'_2+k'_3, 0}\nonumber\\
=\,&2
\sum_{k_1,k_2}
(\xi_{k_1}+\xi_{k_2}+\xi_{-k_1-k_2})
\Omega_{k_1, k_2}^* 
(\Omega_{k_1, k_2}+\Omega_{k_2,-k_1-k_2}+\Omega_{-k_1-k_2,k_1}),
\end{align}
\begin{align}
\left\langle U_2\right\rangle
=\,&
\frac{U_0}{2} 
\sum_{p_1, p_2, p'_1, p'_2} 
\sum_{k_1, k_2} 
\sum_{k'_1, k'_2} 
\Omega_{k_1, k_2}^* 
\Omega_{k'_1, k'_2}
\left\langle 0\right|
A_{k_1, k_2} 
b_{p_1}^{\dagger} b_{p_2}^{\dagger} 
b_{p'_2} b_{p'_1} 
A_{k'_1, k'_2}^{\dagger} 
\left|0\right\rangle 
\delta_{p_1+p_2, p'_1+p'_2}\nonumber\\
=\,&2{U_0} 
\sum_{k_1, k_2,q} 
\Omega_{k_1, k_2}^* 
\left(
2\Omega_{-k_1-k_2,q}
+
\Omega_{q, k_1+k_2-q}
+
2\Omega_{k_2, q}
+
\Omega_{q, -q-k_2}
+
2\Omega_{k_1, q}
+
\Omega_{q, -q-k_1}
\right)
,
\end{align}
\begin{align}
\left\langle U_{\rm eff}^{(2)}\right\rangle
=\,&
\frac{V_2}{2} 
\sum_{k,k',q} 
\sum_{k_1, k_2} 
\sum_{k'_1, k'_2} 
\Omega_{k_1, k_2}^* 
\Omega_{k'_1, k'_2}
\left\langle 0\right|
A_{k_1, k_2}
b_{k+q}^{\dagger} b_{k'-q}^{\dagger} 
b_{k'} b_{k} 
A_{k'_1, k'_2}^{\dagger} 
\left|0\right\rangle 
\nonumber\\
=\,&2V_2
\sum_{k_1, k_2,q} 
\Omega_{k_1, k_2}^*
\left(
2\Omega_{-k_1-k_2,q}
+
\Omega_{q, k_1+k_2-q}
+
2\Omega_{k_2, q}
+
\Omega_{q, -q-k_2}
+
2\Omega_{k_1, q}
+
\Omega_{q, -q-k_1}
\right)
,
\end{align}
and
\begin{align}
\left\langle U_{\rm eff}^{(3)}\right\rangle
=\,&
\frac{V_3}{6} 
\sum_{p_1,p_2,p_3} 
\sum_{p_1',p_2',p_3'} 
\sum_{k_1, k_2} 
\sum_{k'_1, k'_2} 
\Omega_{k_1, k_2}^* 
\Omega_{k'_1, k'_2}
\left\langle 0\right|
A_{k_1, k_2}
b_{p_1'}^{\dagger} 
b_{p_2'}^{\dagger} 
b_{p_3'}^{\dagger} 
b_{p_3}
b_{p_2}
b_{p_1} 
A_{k'_1, k'_2}^{\dagger} 
\left|0\right\rangle 
\nonumber\\
=\,& 
6V_3\sum_{p_1,p_2}
\sum_{k_1,k_2}
\Omega_{k_1, k_2}^* 
\Omega_{p_1, p_2},
\end{align}
respectively.
\end{widetext}

%\bibliography{GYX}

\begin{thebibliography}{62}%
\makeatletter
\providecommand \@ifxundefined [1]{%
 \@ifx{#1\undefined}
}%
\providecommand \@ifnum [1]{%
 \ifnum #1\expandafter \@firstoftwo
 \else \expandafter \@secondoftwo
 \fi
}%
\providecommand \@ifx [1]{%
 \ifx #1\expandafter \@firstoftwo
 \else \expandafter \@secondoftwo
 \fi
}%
\providecommand \natexlab [1]{#1}%
\providecommand \enquote  [1]{``#1''}%
\providecommand \bibnamefont  [1]{#1}%
\providecommand \bibfnamefont [1]{#1}%
\providecommand \citenamefont [1]{#1}%
\providecommand \href@noop [0]{\@secondoftwo}%
\providecommand \href [0]{\begingroup \@sanitize@url \@href}%
\providecommand \@href[1]{\@@startlink{#1}\@@href}%
\providecommand \@@href[1]{\endgroup#1\@@endlink}%
\providecommand \@sanitize@url [0]{\catcode `\\12\catcode `\$12\catcode
  `\&12\catcode `\#12\catcode `\^12\catcode `\_12\catcode `\%12\relax}%
\providecommand \@@startlink[1]{}%
\providecommand \@@endlink[0]{}%
\providecommand \url  [0]{\begingroup\@sanitize@url \@url }%
\providecommand \@url [1]{\endgroup\@href {#1}{\urlprefix }}%
\providecommand \urlprefix  [0]{URL }%
\providecommand \Eprint [0]{\href }%
\providecommand \doibase [0]{https://doi.org/}%
\providecommand \selectlanguage [0]{\@gobble}%
\providecommand \bibinfo  [0]{\@secondoftwo}%
\providecommand \bibfield  [0]{\@secondoftwo}%
\providecommand \translation [1]{[#1]}%
\providecommand \BibitemOpen [0]{}%
\providecommand \bibitemStop [0]{}%
\providecommand \bibitemNoStop [0]{.\EOS\space}%
\providecommand \EOS [0]{\spacefactor3000\relax}%
\providecommand \BibitemShut  [1]{\csname bibitem#1\endcsname}%
\let\auto@bib@innerbib\@empty
%</preamble>
\bibitem [{\citenamefont {Freer}\ \emph {et~al.}(2018)\citenamefont {Freer},
  \citenamefont {Horiuchi}, \citenamefont {Kanada-En'yo}, \citenamefont {Lee},\
  and\ \citenamefont {Mei\ss{}ner}}]{Freer2018Rev.Mod.Phys.90.035004}%
  \BibitemOpen
  \bibfield  {author} {\bibinfo {author} {\bibfnamefont {M.}~\bibnamefont
  {Freer}}, \bibinfo {author} {\bibfnamefont {H.}~\bibnamefont {Horiuchi}},
  \bibinfo {author} {\bibfnamefont {Y.}~\bibnamefont {Kanada-En'yo}}, \bibinfo
  {author} {\bibfnamefont {D.}~\bibnamefont {Lee}},\ and\ \bibinfo {author}
  {\bibfnamefont {U.-G.}\ \bibnamefont {Mei\ss{}ner}},\ }\bibfield  {title}
  {\bibinfo {title} {Microscopic clustering in light nuclei},\ }\href
  {https://doi.org/10.1103/RevModPhys.90.035004} {\bibfield  {journal}
  {\bibinfo  {journal} {Rev. Mod. Phys.}\ }\textbf {\bibinfo {volume} {90}},\
  \bibinfo {pages} {035004} (\bibinfo {year} {2018})}\BibitemShut {NoStop}%
\bibitem [{\citenamefont {Zwierlein}\ \emph {et~al.}(2006)\citenamefont
  {Zwierlein}, \citenamefont {Schirotzek}, \citenamefont {Schunck},\ and\
  \citenamefont {Ketterle}}]{Zwierlein2006Science311.492--496}%
  \BibitemOpen
  \bibfield  {author} {\bibinfo {author} {\bibfnamefont {M.~W.}\ \bibnamefont
  {Zwierlein}}, \bibinfo {author} {\bibfnamefont {A.}~\bibnamefont
  {Schirotzek}}, \bibinfo {author} {\bibfnamefont {C.~H.}\ \bibnamefont
  {Schunck}},\ and\ \bibinfo {author} {\bibfnamefont {W.}~\bibnamefont
  {Ketterle}},\ }\bibfield  {title} {\bibinfo {title} {Fermionic superfluidity
  with imbalanced spin populations},\ }\href
  {https://doi.org/10.1126/science.1122318} {\bibfield  {journal} {\bibinfo
  {journal} {Science}\ }\textbf {\bibinfo {volume} {311}},\ \bibinfo {pages}
  {492} (\bibinfo {year} {2006})},\ \Eprint
  {https://arxiv.org/abs/https://www.science.org/doi/pdf/10.1126/science.1122318}
  {https://www.science.org/doi/pdf/10.1126/science.1122318} \BibitemShut
  {NoStop}%
\bibitem [{\citenamefont {Balents}\ \emph {et~al.}(2020)\citenamefont
  {Balents}, \citenamefont {Dean}, \citenamefont {Efetov},\ and\ \citenamefont
  {Young}}]{Balents2020Nat.Phys.16.725--733}%
  \BibitemOpen
  \bibfield  {author} {\bibinfo {author} {\bibfnamefont {L.}~\bibnamefont
  {Balents}}, \bibinfo {author} {\bibfnamefont {C.~R.}\ \bibnamefont {Dean}},
  \bibinfo {author} {\bibfnamefont {D.~K.}\ \bibnamefont {Efetov}},\ and\
  \bibinfo {author} {\bibfnamefont {A.~F.}\ \bibnamefont {Young}},\ }\bibfield
  {title} {\bibinfo {title} {Superconductivity and strong correlations in
  moiréflat bands},\ }\href {https://doi.org/10.1038/s41567-020-0906-9}
  {\bibfield  {journal} {\bibinfo  {journal} {Nat. Phys.}\ }\textbf {\bibinfo
  {volume} {16}},\ \bibinfo {pages} {725} (\bibinfo {year} {2020})}\BibitemShut
  {NoStop}%
\bibitem [{\citenamefont {Horiuchi}\ \emph {et~al.}(2012)\citenamefont
  {Horiuchi}, \citenamefont {Ikeda},\ and\ \citenamefont
  {Katō}}]{Horiuchi2012Prog.Theor.Phys.Suppl.192.1--238}%
  \BibitemOpen
  \bibfield  {author} {\bibinfo {author} {\bibfnamefont {H.}~\bibnamefont
  {Horiuchi}}, \bibinfo {author} {\bibfnamefont {K.}~\bibnamefont {Ikeda}},\
  and\ \bibinfo {author} {\bibfnamefont {K.}~\bibnamefont {Katō}},\ }\bibfield
   {title} {\bibinfo {title} {{Recent Developments in Nuclear Cluster
  Physics}},\ }\href {https://doi.org/10.1143/PTPS.192.1} {\bibfield  {journal}
  {\bibinfo  {journal} {Prog. Theor. Phys. Suppl.}\ }\textbf {\bibinfo {volume}
  {192}},\ \bibinfo {pages} {1} (\bibinfo {year} {2012})}\BibitemShut {NoStop}%
\bibitem [{\citenamefont {Bai}\ \emph {et~al.}(2019)\citenamefont {Bai},
  \citenamefont {Ren},\ and\ \citenamefont
  {R\"opke}}]{Bai2019Phys.Rev.C99.034305}%
  \BibitemOpen
  \bibfield  {author} {\bibinfo {author} {\bibfnamefont {D.}~\bibnamefont
  {Bai}}, \bibinfo {author} {\bibfnamefont {Z.}~\bibnamefont {Ren}},\ and\
  \bibinfo {author} {\bibfnamefont {G.}~\bibnamefont {R\"opke}},\ }\bibfield
  {title} {\bibinfo {title} {$\ensuremath{\alpha}$ clustering from the quartet
  model},\ }\href {https://doi.org/10.1103/PhysRevC.99.034305} {\bibfield
  {journal} {\bibinfo  {journal} {Phys. Rev. C}\ }\textbf {\bibinfo {volume}
  {99}},\ \bibinfo {pages} {034305} (\bibinfo {year} {2019})}\BibitemShut
  {NoStop}%
\bibitem [{\citenamefont {Itagaki}\ and\ \citenamefont
  {Okabe}(2000)}]{Itagaki2000Phys.Rev.C61.044306}%
  \BibitemOpen
  \bibfield  {author} {\bibinfo {author} {\bibfnamefont {N.}~\bibnamefont
  {Itagaki}}\ and\ \bibinfo {author} {\bibfnamefont {S.}~\bibnamefont
  {Okabe}},\ }\bibfield  {title} {\bibinfo {title} {Molecular orbital
  structures in ${}^{10}\mathrm{Be}$},\ }\href
  {https://doi.org/10.1103/PhysRevC.61.044306} {\bibfield  {journal} {\bibinfo
  {journal} {Phys. Rev. C}\ }\textbf {\bibinfo {volume} {61}},\ \bibinfo
  {pages} {044306} (\bibinfo {year} {2000})}\BibitemShut {NoStop}%
\bibitem [{\citenamefont {Elhatisari}\ \emph {et~al.}(2017)\citenamefont
  {Elhatisari}, \citenamefont {Epelbaum}, \citenamefont {Krebs}, \citenamefont
  {L\"ahde}, \citenamefont {Lee}, \citenamefont {Li}, \citenamefont {Lu},
  \citenamefont {Mei\ss{}ner},\ and\ \citenamefont
  {Rupak}}]{Elhatisari2017Phys.Rev.Lett.119.222505}%
  \BibitemOpen
  \bibfield  {author} {\bibinfo {author} {\bibfnamefont {S.}~\bibnamefont
  {Elhatisari}}, \bibinfo {author} {\bibfnamefont {E.}~\bibnamefont
  {Epelbaum}}, \bibinfo {author} {\bibfnamefont {H.}~\bibnamefont {Krebs}},
  \bibinfo {author} {\bibfnamefont {T.~A.}\ \bibnamefont {L\"ahde}}, \bibinfo
  {author} {\bibfnamefont {D.}~\bibnamefont {Lee}}, \bibinfo {author}
  {\bibfnamefont {N.}~\bibnamefont {Li}}, \bibinfo {author} {\bibfnamefont
  {B.-n.}\ \bibnamefont {Lu}}, \bibinfo {author} {\bibfnamefont {U.-G.}\
  \bibnamefont {Mei\ss{}ner}},\ and\ \bibinfo {author} {\bibfnamefont
  {G.}~\bibnamefont {Rupak}},\ }\bibfield  {title} {\bibinfo {title} {Ab initio
  calculations of the isotopic dependence of nuclear clustering},\ }\href
  {https://doi.org/10.1103/PhysRevLett.119.222505} {\bibfield  {journal}
  {\bibinfo  {journal} {Phys. Rev. Lett.}\ }\textbf {\bibinfo {volume} {119}},\
  \bibinfo {pages} {222505} (\bibinfo {year} {2017})}\BibitemShut {NoStop}%
\bibitem [{\citenamefont {Tanaka}\ \emph {et~al.}(2021)\citenamefont {Tanaka},
  \citenamefont {Yang}, \citenamefont {Typel}, \citenamefont {Adachi},
  \citenamefont {Bai}, \citenamefont {van Beek}, \citenamefont {Beaumel},
  \citenamefont {Fujikawa}, \citenamefont {Han}, \citenamefont {Heil} \emph
  {et~al.}}]{Tanaka2021Science.371.260}%
  \BibitemOpen
  \bibfield  {author} {\bibinfo {author} {\bibfnamefont {J.}~\bibnamefont
  {Tanaka}}, \bibinfo {author} {\bibfnamefont {Z.}~\bibnamefont {Yang}},
  \bibinfo {author} {\bibfnamefont {S.}~\bibnamefont {Typel}}, \bibinfo
  {author} {\bibfnamefont {S.}~\bibnamefont {Adachi}}, \bibinfo {author}
  {\bibfnamefont {S.}~\bibnamefont {Bai}}, \bibinfo {author} {\bibfnamefont
  {P.}~\bibnamefont {van Beek}}, \bibinfo {author} {\bibfnamefont
  {D.}~\bibnamefont {Beaumel}}, \bibinfo {author} {\bibfnamefont
  {Y.}~\bibnamefont {Fujikawa}}, \bibinfo {author} {\bibfnamefont
  {J.}~\bibnamefont {Han}}, \bibinfo {author} {\bibfnamefont {S.}~\bibnamefont
  {Heil}}, \emph {et~al.},\ }\bibfield  {title} {\bibinfo {title} {Formation of
  $\alpha$ clusters in dilute neutron-rich matter},\ } {\bibfield
  {journal} {\bibinfo  {journal} {Science}\ }\textbf {\bibinfo {volume}
  {371}},\ \bibinfo {pages} {260} (\bibinfo {year} {2021})}\BibitemShut
  {NoStop}%
\bibitem [{\citenamefont {Moriya}\ \emph {et~al.}(2021)\citenamefont {Moriya},
  \citenamefont {Tajima}, \citenamefont {Horiuchi}, \citenamefont {Iida},\ and\
  \citenamefont {Nakano}}]{Moriya2021Phys.Rev.C104.065801}%
  \BibitemOpen
  \bibfield  {author} {\bibinfo {author} {\bibfnamefont {H.}~\bibnamefont
  {Moriya}}, \bibinfo {author} {\bibfnamefont {H.}~\bibnamefont {Tajima}},
  \bibinfo {author} {\bibfnamefont {W.}~\bibnamefont {Horiuchi}}, \bibinfo
  {author} {\bibfnamefont {K.}~\bibnamefont {Iida}},\ and\ \bibinfo {author}
  {\bibfnamefont {E.}~\bibnamefont {Nakano}},\ }\bibfield  {title} {\bibinfo
  {title} {Binding two and three $\ensuremath{\alpha}$ particles in cold
  neutron matter},\ }\href {https://doi.org/10.1103/PhysRevC.104.065801}
  {\bibfield  {journal} {\bibinfo  {journal} {Phys. Rev. C}\ }\textbf {\bibinfo
  {volume} {104}},\ \bibinfo {pages} {065801} (\bibinfo {year}
  {2021})}\BibitemShut {NoStop}%
\bibitem [{\citenamefont {Tajima}\ \emph {et~al.}(2023)\citenamefont {Tajima},
  \citenamefont {Moriya}, \citenamefont {Horiuchi}, \citenamefont {Nakano},\
  and\ \citenamefont {Iida}}]{tajima2023polaronic}%
  \BibitemOpen
  \bibfield  {author} {\bibinfo {author} {\bibfnamefont {H.}~\bibnamefont
  {Tajima}}, \bibinfo {author} {\bibfnamefont {H.}~\bibnamefont {Moriya}},
  \bibinfo {author} {\bibfnamefont {W.}~\bibnamefont {Horiuchi}}, \bibinfo
  {author} {\bibfnamefont {E.}~\bibnamefont {Nakano}},\ and\ \bibinfo {author}
  {\bibfnamefont {K.}~\bibnamefont {Iida}},\ } {\bibinfo {title}
  {Polaronic proton and diproton clustering in neutron-rich matter}} (\bibinfo
  {year} {2023}),\ \Eprint {https://arxiv.org/abs/2304.00535} {arXiv:2304.00535
  [nucl-th]} \BibitemShut {NoStop}%
\bibitem [{\citenamefont {Chevy}\ and\ \citenamefont
  {Mora}(2010)}]{Chevy2010Rep.Prog.Phys.73.112401}%
  \BibitemOpen
  \bibfield  {author} {\bibinfo {author} {\bibfnamefont {F.}~\bibnamefont
  {Chevy}}\ and\ \bibinfo {author} {\bibfnamefont {C.}~\bibnamefont {Mora}},\
  }\bibfield  {title} {\bibinfo {title} {Ultra-cold polarized fermi gases},\
  }\href {https://doi.org/10.1088/0034-4885/73/11/112401} {\bibfield  {journal}
  {\bibinfo  {journal} {Rep. Prog. Phys.}\ }\textbf {\bibinfo {volume} {73}},\
  \bibinfo {pages} {112401} (\bibinfo {year} {2010})}\BibitemShut {NoStop}%
\bibitem [{\citenamefont {Schmidt}\ \emph {et~al.}(2018)\citenamefont
  {Schmidt}, \citenamefont {Knap}, \citenamefont {Ivanov}, \citenamefont {You},
  \citenamefont {Cetina},\ and\ \citenamefont
  {Demler}}]{Schmidt2018Rep.Prog.Phys.81.024401}%
  \BibitemOpen
  \bibfield  {author} {\bibinfo {author} {\bibfnamefont {R.}~\bibnamefont
  {Schmidt}}, \bibinfo {author} {\bibfnamefont {M.}~\bibnamefont {Knap}},
  \bibinfo {author} {\bibfnamefont {D.~A.}\ \bibnamefont {Ivanov}}, \bibinfo
  {author} {\bibfnamefont {J.-S.}\ \bibnamefont {You}}, \bibinfo {author}
  {\bibfnamefont {M.}~\bibnamefont {Cetina}},\ and\ \bibinfo {author}
  {\bibfnamefont {E.}~\bibnamefont {Demler}},\ }\bibfield  {title} {\bibinfo
  {title} {Universal many-body response of heavy impurities coupled to a fermi
  sea: a review of recent progress},\ }\href
  {https://doi.org/10.1088/1361-6633/aa9593} {\bibfield  {journal} {\bibinfo
  {journal} {Rep. Prog. Phys.}\ }\textbf {\bibinfo {volume} {81}},\ \bibinfo
  {pages} {024401} (\bibinfo {year} {2018})}\BibitemShut {NoStop}%
\bibitem [{\citenamefont {Landau}(1933)}]{Landau1933Phys.Z.Sowjetunion3.664}%
  \BibitemOpen
  \bibfield  {author} {\bibinfo {author} {\bibfnamefont {L.}~\bibnamefont
  {Landau}},\ }\bibfield  {title} {\bibinfo {title} {Electron motion in crystal
  lattices},\ } {\bibfield  {journal} {\bibinfo  {journal} {Phys.
  Z. Sowjetunion}\ }\textbf {\bibinfo {volume} {3}},\ \bibinfo {pages} {664}
  (\bibinfo {year} {1933})}\BibitemShut {NoStop}%
\bibitem [{\citenamefont {Landau}\ and\ \citenamefont
  {Pekar}(1948)}]{landau1948effective}%
  \BibitemOpen
  \bibfield  {author} {\bibinfo {author} {\bibfnamefont {L.}~\bibnamefont
  {Landau}}\ and\ \bibinfo {author} {\bibfnamefont {S.}~\bibnamefont {Pekar}},\
  }\bibfield  {title} {\bibinfo {title} {Effective mass of a polaron},\
  } {\bibfield  {journal} {\bibinfo  {journal} {Zh. Eksp. Teor.
  Fiz.}\ }\textbf {\bibinfo {volume} {18}},\ \bibinfo {pages} {419} (\bibinfo
  {year} {1948})}\BibitemShut {NoStop}%
\bibitem [{\citenamefont {Massignan}\ \emph {et~al.}(2014)\citenamefont
  {Massignan}, \citenamefont {Zaccanti},\ and\ \citenamefont
  {Bruun}}]{Massignan2014Rep.Prog.Phys.77.034401}%
  \BibitemOpen
  \bibfield  {author} {\bibinfo {author} {\bibfnamefont {P.}~\bibnamefont
  {Massignan}}, \bibinfo {author} {\bibfnamefont {M.}~\bibnamefont
  {Zaccanti}},\ and\ \bibinfo {author} {\bibfnamefont {G.~M.}\ \bibnamefont
  {Bruun}},\ }\bibfield  {title} {\bibinfo {title} {Polarons, dressed molecules
  and itinerant ferromagnetism in ultracold fermi gases},\ }\href
  {https://doi.org/10.1088/0034-4885/77/3/034401} {\bibfield  {journal}
  {\bibinfo  {journal} {Rep. Prog. Phys.}\ }\textbf {\bibinfo {volume} {77}},\
  \bibinfo {pages} {034401} (\bibinfo {year} {2014})}\BibitemShut {NoStop}%
\bibitem [{\citenamefont {Fisher}\ \emph {et~al.}(1989)\citenamefont {Fisher},
  \citenamefont {Hayes},\ and\ \citenamefont
  {Wallace}}]{Fisher1989J.Phys.:Condens.Matter1.5567}%
  \BibitemOpen
  \bibfield  {author} {\bibinfo {author} {\bibfnamefont {A.~J.}\ \bibnamefont
  {Fisher}}, \bibinfo {author} {\bibfnamefont {W.}~\bibnamefont {Hayes}},\ and\
  \bibinfo {author} {\bibfnamefont {D.~S.}\ \bibnamefont {Wallace}},\
  }\bibfield  {title} {\bibinfo {title} {Polarons and solitons},\ }\href
  {https://doi.org/10.1088/0953-8984/1/33/001} {\bibfield  {journal} {\bibinfo
  {journal} {J. Phys.: Condens. Matter}\ }\textbf {\bibinfo {volume} {1}},\
  \bibinfo {pages} {5567} (\bibinfo {year} {1989})}\BibitemShut {NoStop}%
\bibitem [{\citenamefont {Alexandrov}\ and\ \citenamefont
  {Mott}(1994)}]{Alexandrov1994Rep.Prog.Phys.57.1197}%
  \BibitemOpen
  \bibfield  {author} {\bibinfo {author} {\bibfnamefont {A.~S.}\ \bibnamefont
  {Alexandrov}}\ and\ \bibinfo {author} {\bibfnamefont {N.~F.}\ \bibnamefont
  {Mott}},\ }\bibfield  {title} {\bibinfo {title} {Bipolarons},\ }\href
  {https://doi.org/10.1088/0034-4885/57/12/001} {\bibfield  {journal} {\bibinfo
   {journal} {Rep. Prog. Phys.}\ }\textbf {\bibinfo {volume} {57}},\ \bibinfo
  {pages} {1197} (\bibinfo {year} {1994})}\BibitemShut {NoStop}%
\bibitem [{\citenamefont {Rice}\ \emph {et~al.}(1986)\citenamefont {Rice},
  \citenamefont {Phillpot}, \citenamefont {Bishop},\ and\ \citenamefont
  {Campbell}}]{Rice1986Phys.Rev.B34.4139--4149}%
  \BibitemOpen
  \bibfield  {author} {\bibinfo {author} {\bibfnamefont {M.~J.}\ \bibnamefont
  {Rice}}, \bibinfo {author} {\bibfnamefont {S.~R.}\ \bibnamefont {Phillpot}},
  \bibinfo {author} {\bibfnamefont {A.~R.}\ \bibnamefont {Bishop}},\ and\
  \bibinfo {author} {\bibfnamefont {D.~K.}\ \bibnamefont {Campbell}},\
  }\bibfield  {title} {\bibinfo {title} {Solitons, polarons, and phonons in the
  infinite polyyne chain},\ }\href {https://doi.org/10.1103/PhysRevB.34.4139}
  {\bibfield  {journal} {\bibinfo  {journal} {Phys. Rev. B}\ }\textbf {\bibinfo
  {volume} {34}},\ \bibinfo {pages} {4139} (\bibinfo {year}
  {1986})}\BibitemShut {NoStop}%
\bibitem [{\citenamefont {King}\ \emph {et~al.}(2015)\citenamefont {King},
  \citenamefont {Llobet},\ and\ \citenamefont
  {Garcia-Martin}}]{King2015Phys.Rev.B91.024412}%
  \BibitemOpen
  \bibfield  {author} {\bibinfo {author} {\bibfnamefont {G.}~\bibnamefont
  {King}}, \bibinfo {author} {\bibfnamefont {A.}~\bibnamefont {Llobet}},\ and\
  \bibinfo {author} {\bibfnamefont {S.}~\bibnamefont {Garcia-Martin}},\
  }\bibfield  {title} {\bibinfo {title} {Magnetic properties and magnetic
  structures of $\mathrm{TbBaM}{\mathrm{n}}_{2}{\mathrm{o}}_{5.75}$: Possible
  observation of unconventional polaron trimers},\ }\href
  {https://doi.org/10.1103/PhysRevB.91.024412} {\bibfield  {journal} {\bibinfo
  {journal} {Phys. Rev. B}\ }\textbf {\bibinfo {volume} {91}},\ \bibinfo
  {pages} {024412} (\bibinfo {year} {2015})}\BibitemShut {NoStop}%
\bibitem [{\citenamefont {Sous}\ \emph {et~al.}(2017)\citenamefont {Sous},
  \citenamefont {Berciu},\ and\ \citenamefont
  {Krems}}]{Sous2017Phys.Rev.A96.063619}%
  \BibitemOpen
  \bibfield  {author} {\bibinfo {author} {\bibfnamefont {J.}~\bibnamefont
  {Sous}}, \bibinfo {author} {\bibfnamefont {M.}~\bibnamefont {Berciu}},\ and\
  \bibinfo {author} {\bibfnamefont {R.~V.}\ \bibnamefont {Krems}},\ }\bibfield
  {title} {\bibinfo {title} {Bipolarons bound by repulsive phonon-mediated
  interactions},\ }\href {https://doi.org/10.1103/PhysRevA.96.063619}
  {\bibfield  {journal} {\bibinfo  {journal} {Phys. Rev. A}\ }\textbf {\bibinfo
  {volume} {96}},\ \bibinfo {pages} {063619} (\bibinfo {year}
  {2017})}\BibitemShut {NoStop}%
\bibitem [{\citenamefont {Camacho-Guardian}\ \emph {et~al.}(2018)\citenamefont
  {Camacho-Guardian}, \citenamefont {Pe\~na Ardila}, \citenamefont {Pohl},\
  and\ \citenamefont {Bruun}}]{Camacho-Guardian2018Phys.Rev.Lett.121.013401}%
  \BibitemOpen
  \bibfield  {author} {\bibinfo {author} {\bibfnamefont {A.}~\bibnamefont
  {Camacho-Guardian}}, \bibinfo {author} {\bibfnamefont {L.~A.}\ \bibnamefont
  {Pe\~na Ardila}}, \bibinfo {author} {\bibfnamefont {T.}~\bibnamefont
  {Pohl}},\ and\ \bibinfo {author} {\bibfnamefont {G.~M.}\ \bibnamefont
  {Bruun}},\ }\bibfield  {title} {\bibinfo {title} {Bipolarons in a
  bose-einstein condensate},\ }\href
  {https://doi.org/10.1103/PhysRevLett.121.013401} {\bibfield  {journal}
  {\bibinfo  {journal} {Phys. Rev. Lett.}\ }\textbf {\bibinfo {volume} {121}},\
  \bibinfo {pages} {013401} (\bibinfo {year} {2018})}\BibitemShut {NoStop}%
\bibitem [{\citenamefont {Eismann}\ \emph {et~al.}(2016)\citenamefont
  {Eismann}, \citenamefont {Khaykovich}, \citenamefont {Laurent}, \citenamefont
  {Ferrier-Barbut}, \citenamefont {Rem}, \citenamefont {Grier}, \citenamefont
  {Delehaye}, \citenamefont {Chevy}, \citenamefont {Salomon}, \citenamefont
  {Ha},\ and\ \citenamefont {Chin}}]{Eismann2016Phys.Rev.X6.021025}%
  \BibitemOpen
  \bibfield  {author} {\bibinfo {author} {\bibfnamefont {U.}~\bibnamefont
  {Eismann}}, \bibinfo {author} {\bibfnamefont {L.}~\bibnamefont {Khaykovich}},
  \bibinfo {author} {\bibfnamefont {S.}~\bibnamefont {Laurent}}, \bibinfo
  {author} {\bibfnamefont {I.}~\bibnamefont {Ferrier-Barbut}}, \bibinfo
  {author} {\bibfnamefont {B.~S.}\ \bibnamefont {Rem}}, \bibinfo {author}
  {\bibfnamefont {A.~T.}\ \bibnamefont {Grier}}, \bibinfo {author}
  {\bibfnamefont {M.}~\bibnamefont {Delehaye}}, \bibinfo {author}
  {\bibfnamefont {F.}~\bibnamefont {Chevy}}, \bibinfo {author} {\bibfnamefont
  {C.}~\bibnamefont {Salomon}}, \bibinfo {author} {\bibfnamefont {L.-C.}\
  \bibnamefont {Ha}},\ and\ \bibinfo {author} {\bibfnamefont {C.}~\bibnamefont
  {Chin}},\ }\bibfield  {title} {\bibinfo {title} {Universal loss dynamics in a
  unitary bose gas},\ }\href {https://doi.org/10.1103/PhysRevX.6.021025}
  {\bibfield  {journal} {\bibinfo  {journal} {Phys. Rev. X}\ }\textbf {\bibinfo
  {volume} {6}},\ \bibinfo {pages} {021025} (\bibinfo {year}
  {2016})}\BibitemShut {NoStop}%
\bibitem [{\citenamefont {Barontini}\ \emph {et~al.}(2009)\citenamefont
  {Barontini}, \citenamefont {Weber}, \citenamefont {Rabatti}, \citenamefont
  {Catani}, \citenamefont {Thalhammer}, \citenamefont {Inguscio},\ and\
  \citenamefont {Minardi}}]{Barontini2009Phys.Rev.Lett.103.043201}%
  \BibitemOpen
  \bibfield  {author} {\bibinfo {author} {\bibfnamefont {G.}~\bibnamefont
  {Barontini}}, \bibinfo {author} {\bibfnamefont {C.}~\bibnamefont {Weber}},
  \bibinfo {author} {\bibfnamefont {F.}~\bibnamefont {Rabatti}}, \bibinfo
  {author} {\bibfnamefont {J.}~\bibnamefont {Catani}}, \bibinfo {author}
  {\bibfnamefont {G.}~\bibnamefont {Thalhammer}}, \bibinfo {author}
  {\bibfnamefont {M.}~\bibnamefont {Inguscio}},\ and\ \bibinfo {author}
  {\bibfnamefont {F.}~\bibnamefont {Minardi}},\ }\bibfield  {title} {\bibinfo
  {title} {Observation of heteronuclear atomic efimov resonances},\ }\href
  {https://doi.org/10.1103/PhysRevLett.103.043201} {\bibfield  {journal}
  {\bibinfo  {journal} {Phys. Rev. Lett.}\ }\textbf {\bibinfo {volume} {103}},\
  \bibinfo {pages} {043201} (\bibinfo {year} {2009})}\BibitemShut {NoStop}%
\bibitem [{\citenamefont {Niemann}\ and\ \citenamefont
  {Hammer}(2012)}]{Niemann2012Phys.Rev.A86.013628}%
  \BibitemOpen
  \bibfield  {author} {\bibinfo {author} {\bibfnamefont {P.}~\bibnamefont
  {Niemann}}\ and\ \bibinfo {author} {\bibfnamefont {H.-W.}\ \bibnamefont
  {Hammer}},\ }\bibfield  {title} {\bibinfo {title} {Pauli-blocking effects and
  cooper triples in three-component fermi gases},\ }\href
  {https://doi.org/10.1103/PhysRevA.86.013628} {\bibfield  {journal} {\bibinfo
  {journal} {Phys. Rev. A}\ }\textbf {\bibinfo {volume} {86}},\ \bibinfo
  {pages} {013628} (\bibinfo {year} {2012})}\BibitemShut {NoStop}%
\bibitem [{\citenamefont {Nygaard}\ and\ \citenamefont
  {Zinner}(2014)}]{nygaard2014efimov}%
  \BibitemOpen
  \bibfield  {author} {\bibinfo {author} {\bibfnamefont {N.~G.}\ \bibnamefont
  {Nygaard}}\ and\ \bibinfo {author} {\bibfnamefont {N.~T.}\ \bibnamefont
  {Zinner}},\ }\bibfield  {title} {\bibinfo {title} {Efimov three-body states
  on top of a fermi sea},\ }{\bibfield  {journal} {\bibinfo
  {journal} {New Journal of Physics}\ }\textbf {\bibinfo {volume} {16}},\
  \bibinfo {pages} {023026} (\bibinfo {year} {2014})}\BibitemShut {NoStop}%
\bibitem [{\citenamefont {Kirk}\ and\ \citenamefont
  {Parish}(2017)}]{Kirk2017Phys.Rev.A96.053614}%
  \BibitemOpen
  \bibfield  {author} {\bibinfo {author} {\bibfnamefont {T.}~\bibnamefont
  {Kirk}}\ and\ \bibinfo {author} {\bibfnamefont {M.~M.}\ \bibnamefont
  {Parish}},\ }\bibfield  {title} {\bibinfo {title} {Three-body correlations in
  a two-dimensional su(3) fermi gas},\ }\href
  {https://doi.org/10.1103/PhysRevA.96.053614} {\bibfield  {journal} {\bibinfo
  {journal} {Phys. Rev. A}\ }\textbf {\bibinfo {volume} {96}},\ \bibinfo
  {pages} {053614} (\bibinfo {year} {2017})}\BibitemShut {NoStop}%
\bibitem [{\citenamefont {Sun}\ and\ \citenamefont
  {Cui}(2019)}]{Sun2019PhysRevA.99.060701}%
  \BibitemOpen
  \bibfield  {author} {\bibinfo {author} {\bibfnamefont {M.}~\bibnamefont
  {Sun}}\ and\ \bibinfo {author} {\bibfnamefont {X.}~\bibnamefont {Cui}},\
  }\bibfield  {title} {\bibinfo {title} {Efimov physics in the presence of a
  fermi sea},\ }\href {https://doi.org/10.1103/PhysRevA.99.060701} {\bibfield
  {journal} {\bibinfo  {journal} {Phys. Rev. A}\ }\textbf {\bibinfo {volume}
  {99}},\ \bibinfo {pages} {060701} (\bibinfo {year} {2019})}\BibitemShut
  {NoStop}%
\bibitem [{\citenamefont {Tajima}\ and\ \citenamefont
  {Naidon}(2019)}]{tajima2019quantum}%
  \BibitemOpen
  \bibfield  {author} {\bibinfo {author} {\bibfnamefont {H.}~\bibnamefont
  {Tajima}}\ and\ \bibinfo {author} {\bibfnamefont {P.}~\bibnamefont
  {Naidon}},\ }\bibfield  {title} {\bibinfo {title} {Quantum chromodynamics
  (qcd)-like phase diagram with efimov trimers and cooper pairs in resonantly
  interacting su (3) fermi gases},\ } {\bibfield  {journal}
  {\bibinfo  {journal} {New Journal of Physics}\ }\textbf {\bibinfo {volume}
  {21}},\ \bibinfo {pages} {073051} (\bibinfo {year} {2019})}\BibitemShut
  {NoStop}%
\bibitem [{\citenamefont {Tajima}\ \emph
  {et~al.}(2021{\natexlab{a}})\citenamefont {Tajima}, \citenamefont {Tsutsui},
  \citenamefont {Doi},\ and\ \citenamefont
  {Iida}}]{Tajima2021Phys.Rev.A104.053328}%
  \BibitemOpen
  \bibfield  {author} {\bibinfo {author} {\bibfnamefont {H.}~\bibnamefont
  {Tajima}}, \bibinfo {author} {\bibfnamefont {S.}~\bibnamefont {Tsutsui}},
  \bibinfo {author} {\bibfnamefont {T.~M.}\ \bibnamefont {Doi}},\ and\ \bibinfo
  {author} {\bibfnamefont {K.}~\bibnamefont {Iida}},\ }\bibfield  {title}
  {\bibinfo {title} {Three-body crossover from a cooper triple to a bound
  trimer state in three-component fermi gases near a triatomic resonance},\
  }\href {https://doi.org/10.1103/PhysRevA.104.053328} {\bibfield  {journal}
  {\bibinfo  {journal} {Phys. Rev. A}\ }\textbf {\bibinfo {volume} {104}},\
  \bibinfo {pages} {053328} (\bibinfo {year} {2021}{\natexlab{a}})}\BibitemShut
  {NoStop}%
\bibitem [{\citenamefont {Tajima}\ \emph {et~al.}(2022)\citenamefont {Tajima},
  \citenamefont {Tsutsui}, \citenamefont {Doi},\ and\ \citenamefont
  {Iida}}]{Tajima2022Phys.Rev.Research4.L012021}%
  \BibitemOpen
  \bibfield  {author} {\bibinfo {author} {\bibfnamefont {H.}~\bibnamefont
  {Tajima}}, \bibinfo {author} {\bibfnamefont {S.}~\bibnamefont {Tsutsui}},
  \bibinfo {author} {\bibfnamefont {T.~M.}\ \bibnamefont {Doi}},\ and\ \bibinfo
  {author} {\bibfnamefont {K.}~\bibnamefont {Iida}},\ }\bibfield  {title}
  {\bibinfo {title} {Cooper triples in attractive three-component fermions:
  Implication for hadron-quark crossover},\ }\href
  {https://doi.org/10.1103/PhysRevResearch.4.L012021} {\bibfield  {journal}
  {\bibinfo  {journal} {Phys. Rev. Research}\ }\textbf {\bibinfo {volume}
  {4}},\ \bibinfo {pages} {L012021} (\bibinfo {year} {2022})}\BibitemShut
  {NoStop}%
\bibitem [{\citenamefont {Guo}\ and\ \citenamefont
  {Tajima}(2022)}]{Guo2022Phys.Rev.A106.043310}%
  \BibitemOpen
  \bibfield  {author} {\bibinfo {author} {\bibfnamefont {Y.}~\bibnamefont
  {Guo}}\ and\ \bibinfo {author} {\bibfnamefont {H.}~\bibnamefont {Tajima}},\
  }\bibfield  {title} {\bibinfo {title} {Stability against three-body
  clustering in one-dimensional spinless $p$-wave fermions},\ }\href
  {https://doi.org/10.1103/PhysRevA.106.043310} {\bibfield  {journal} {\bibinfo
   {journal} {Phys. Rev. A}\ }\textbf {\bibinfo {volume} {106}},\ \bibinfo
  {pages} {043310} (\bibinfo {year} {2022})}\BibitemShut {NoStop}%
\bibitem [{\citenamefont {Musolino}\ \emph {et~al.}(2022)\citenamefont
  {Musolino}, \citenamefont {Kurkjian}, \citenamefont {Van~Regemortel},
  \citenamefont {Wouters}, \citenamefont {Kokkelmans},\ and\ \citenamefont
  {Colussi}}]{Musolino2022Phys.Rev.Lett.128.020401}%
  \BibitemOpen
  \bibfield  {author} {\bibinfo {author} {\bibfnamefont {S.}~\bibnamefont
  {Musolino}}, \bibinfo {author} {\bibfnamefont {H.}~\bibnamefont {Kurkjian}},
  \bibinfo {author} {\bibfnamefont {M.}~\bibnamefont {Van~Regemortel}},
  \bibinfo {author} {\bibfnamefont {M.}~\bibnamefont {Wouters}}, \bibinfo
  {author} {\bibfnamefont {S.~J. J. M.~F.}\ \bibnamefont {Kokkelmans}},\ and\
  \bibinfo {author} {\bibfnamefont {V.~E.}\ \bibnamefont {Colussi}},\
  }\bibfield  {title} {\bibinfo {title} {Bose-einstein condensation of
  efimovian triples in the unitary bose gas},\ }\href
  {https://doi.org/10.1103/PhysRevLett.128.020401} {\bibfield  {journal}
  {\bibinfo  {journal} {Phys. Rev. Lett.}\ }\textbf {\bibinfo {volume} {128}},\
  \bibinfo {pages} {020401} (\bibinfo {year} {2022})}\BibitemShut {NoStop}%
\bibitem [{\citenamefont {Guo}\ and\ \citenamefont
  {Tajima}(2023)}]{Guo2023Phys.Rev.B107.024511}%
  \BibitemOpen
  \bibfield  {author} {\bibinfo {author} {\bibfnamefont {Y.}~\bibnamefont
  {Guo}}\ and\ \bibinfo {author} {\bibfnamefont {H.}~\bibnamefont {Tajima}},\
  }\bibfield  {title} {\bibinfo {title} {Competition between pairing and
  tripling in one-dimensional fermions with coexistent $s$- and $p$-wave
  interactions},\ }\href {https://doi.org/10.1103/PhysRevB.107.024511}
  {\bibfield  {journal} {\bibinfo  {journal} {Phys. Rev. B}\ }\textbf {\bibinfo
  {volume} {107}},\ \bibinfo {pages} {024511} (\bibinfo {year}
  {2023})}\BibitemShut {NoStop}%
\bibitem [{\citenamefont {Burt}\ \emph {et~al.}(1997)\citenamefont {Burt},
  \citenamefont {Ghrist}, \citenamefont {Myatt}, \citenamefont {Holland},
  \citenamefont {Cornell},\ and\ \citenamefont
  {Wieman}}]{Burt1997Phys.Rev.Lett.79.337--340}%
  \BibitemOpen
  \bibfield  {author} {\bibinfo {author} {\bibfnamefont {E.~A.}\ \bibnamefont
  {Burt}}, \bibinfo {author} {\bibfnamefont {R.~W.}\ \bibnamefont {Ghrist}},
  \bibinfo {author} {\bibfnamefont {C.~J.}\ \bibnamefont {Myatt}}, \bibinfo
  {author} {\bibfnamefont {M.~J.}\ \bibnamefont {Holland}}, \bibinfo {author}
  {\bibfnamefont {E.~A.}\ \bibnamefont {Cornell}},\ and\ \bibinfo {author}
  {\bibfnamefont {C.~E.}\ \bibnamefont {Wieman}},\ }\bibfield  {title}
  {\bibinfo {title} {Coherence, correlations, and collisions: What one learns
  about bose-einstein condensates from their decay},\ }\href
  {https://doi.org/10.1103/PhysRevLett.79.337} {\bibfield  {journal} {\bibinfo
  {journal} {Phys. Rev. Lett.}\ }\textbf {\bibinfo {volume} {79}},\ \bibinfo
  {pages} {337} (\bibinfo {year} {1997})}\BibitemShut {NoStop}%
\bibitem [{\citenamefont {Haller}\ \emph {et~al.}(2011)\citenamefont {Haller},
  \citenamefont {Rabie}, \citenamefont {Mark}, \citenamefont {Danzl},
  \citenamefont {Hart}, \citenamefont {Lauber}, \citenamefont {Pupillo},\ and\
  \citenamefont {N\"agerl}}]{Haller2011Phys.Rev.Lett.107.230404}%
  \BibitemOpen
  \bibfield  {author} {\bibinfo {author} {\bibfnamefont {E.}~\bibnamefont
  {Haller}}, \bibinfo {author} {\bibfnamefont {M.}~\bibnamefont {Rabie}},
  \bibinfo {author} {\bibfnamefont {M.~J.}\ \bibnamefont {Mark}}, \bibinfo
  {author} {\bibfnamefont {J.~G.}\ \bibnamefont {Danzl}}, \bibinfo {author}
  {\bibfnamefont {R.}~\bibnamefont {Hart}}, \bibinfo {author} {\bibfnamefont
  {K.}~\bibnamefont {Lauber}}, \bibinfo {author} {\bibfnamefont
  {G.}~\bibnamefont {Pupillo}},\ and\ \bibinfo {author} {\bibfnamefont {H.-C.}\
  \bibnamefont {N\"agerl}},\ }\bibfield  {title} {\bibinfo {title} {Three-body
  correlation functions and recombination rates for bosons in three dimensions
  and one dimension},\ }\href {https://doi.org/10.1103/PhysRevLett.107.230404}
  {\bibfield  {journal} {\bibinfo  {journal} {Phys. Rev. Lett.}\ }\textbf
  {\bibinfo {volume} {107}},\ \bibinfo {pages} {230404} (\bibinfo {year}
  {2011})}\BibitemShut {NoStop}%
\bibitem [{\citenamefont {Ottenstein}\ \emph {et~al.}(2008)\citenamefont
  {Ottenstein}, \citenamefont {Lompe}, \citenamefont {Kohnen}, \citenamefont
  {Wenz},\ and\ \citenamefont
  {Jochim}}]{Ottenstein2008Phys.Rev.Lett.101.203202}%
  \BibitemOpen
  \bibfield  {author} {\bibinfo {author} {\bibfnamefont {T.~B.}\ \bibnamefont
  {Ottenstein}}, \bibinfo {author} {\bibfnamefont {T.}~\bibnamefont {Lompe}},
  \bibinfo {author} {\bibfnamefont {M.}~\bibnamefont {Kohnen}}, \bibinfo
  {author} {\bibfnamefont {A.~N.}\ \bibnamefont {Wenz}},\ and\ \bibinfo
  {author} {\bibfnamefont {S.}~\bibnamefont {Jochim}},\ }\bibfield  {title}
  {\bibinfo {title} {Collisional stability of a three-component degenerate
  fermi gas},\ }\href {https://doi.org/10.1103/PhysRevLett.101.203202}
  {\bibfield  {journal} {\bibinfo  {journal} {Phys. Rev. Lett.}\ }\textbf
  {\bibinfo {volume} {101}},\ \bibinfo {pages} {203202} (\bibinfo {year}
  {2008})}\BibitemShut {NoStop}%
\bibitem [{\citenamefont {Huckans}\ \emph {et~al.}(2009)\citenamefont
  {Huckans}, \citenamefont {Williams}, \citenamefont {Hazlett}, \citenamefont
  {Stites},\ and\ \citenamefont
  {O'Hara}}]{Huckans2009Phys.Rev.Lett.102.165302}%
  \BibitemOpen
  \bibfield  {author} {\bibinfo {author} {\bibfnamefont {J.~H.}\ \bibnamefont
  {Huckans}}, \bibinfo {author} {\bibfnamefont {J.~R.}\ \bibnamefont
  {Williams}}, \bibinfo {author} {\bibfnamefont {E.~L.}\ \bibnamefont
  {Hazlett}}, \bibinfo {author} {\bibfnamefont {R.~W.}\ \bibnamefont
  {Stites}},\ and\ \bibinfo {author} {\bibfnamefont {K.~M.}\ \bibnamefont
  {O'Hara}},\ }\bibfield  {title} {\bibinfo {title} {Three-body recombination
  in a three-state fermi gas with widely tunable interactions},\ }\href
  {https://doi.org/10.1103/PhysRevLett.102.165302} {\bibfield  {journal}
  {\bibinfo  {journal} {Phys. Rev. Lett.}\ }\textbf {\bibinfo {volume} {102}},\
  \bibinfo {pages} {165302} (\bibinfo {year} {2009})}\BibitemShut {NoStop}%
\bibitem [{\citenamefont {Nishida}(2009)}]{PhysRevA.79.013629}%
  \BibitemOpen
  \bibfield  {author} {\bibinfo {author} {\bibfnamefont {Y.}~\bibnamefont
  {Nishida}},\ }\bibfield  {title} {\bibinfo {title} {Casimir interaction among
  heavy fermions in the bcs-bec crossover},\ }\href
  {https://doi.org/10.1103/PhysRevA.79.013629} {\bibfield  {journal} {\bibinfo
  {journal} {Phys. Rev. A}\ }\textbf {\bibinfo {volume} {79}},\ \bibinfo
  {pages} {013629} (\bibinfo {year} {2009})}\BibitemShut {NoStop}%
\bibitem [{\citenamefont {Enss}\ \emph {et~al.}(2020)\citenamefont {Enss},
  \citenamefont {Tran}, \citenamefont {Rautenberg}, \citenamefont {Gerken},
  \citenamefont {Lippi}, \citenamefont {Drescher}, \citenamefont {Zhu},
  \citenamefont {Weidem\"uller},\ and\ \citenamefont
  {Salmhofer}}]{PhysRevA.102.063321}%
  \BibitemOpen
  \bibfield  {author} {\bibinfo {author} {\bibfnamefont {T.}~\bibnamefont
  {Enss}}, \bibinfo {author} {\bibfnamefont {B.}~\bibnamefont {Tran}}, \bibinfo
  {author} {\bibfnamefont {M.}~\bibnamefont {Rautenberg}}, \bibinfo {author}
  {\bibfnamefont {M.}~\bibnamefont {Gerken}}, \bibinfo {author} {\bibfnamefont
  {E.}~\bibnamefont {Lippi}}, \bibinfo {author} {\bibfnamefont
  {M.}~\bibnamefont {Drescher}}, \bibinfo {author} {\bibfnamefont
  {B.}~\bibnamefont {Zhu}}, \bibinfo {author} {\bibfnamefont {M.}~\bibnamefont
  {Weidem\"uller}},\ and\ \bibinfo {author} {\bibfnamefont {M.}~\bibnamefont
  {Salmhofer}},\ }\bibfield  {title} {\bibinfo {title} {Scattering of two heavy
  fermi polarons: Resonances and quasibound states},\ }\href
  {https://doi.org/10.1103/PhysRevA.102.063321} {\bibfield  {journal} {\bibinfo
   {journal} {Phys. Rev. A}\ }\textbf {\bibinfo {volume} {102}},\ \bibinfo
  {pages} {063321} (\bibinfo {year} {2020})}\BibitemShut {NoStop}%
\bibitem [{\citenamefont {Bougas}\ \emph {et~al.}(2021)\citenamefont {Bougas},
  \citenamefont {Mistakidis}, \citenamefont {Giannakeas},\ and\ \citenamefont
  {Schmelcher}}]{Bougas2021New.J.Phys.23.093022}%
  \BibitemOpen
  \bibfield  {author} {\bibinfo {author} {\bibfnamefont {G.}~\bibnamefont
  {Bougas}}, \bibinfo {author} {\bibfnamefont {S.~I.}\ \bibnamefont
  {Mistakidis}}, \bibinfo {author} {\bibfnamefont {P.}~\bibnamefont
  {Giannakeas}},\ and\ \bibinfo {author} {\bibfnamefont {P.}~\bibnamefont
  {Schmelcher}},\ }\bibfield  {title} {\bibinfo {title} {Few-body correlations
  in two-dimensional bose and fermi ultracold mixtures},\ }\href
  {https://doi.org/10.1088/1367-2630/ac0e56} {\bibfield  {journal} {\bibinfo
  {journal} {New Journal of Physics}\ }\textbf {\bibinfo {volume} {23}},\
  \bibinfo {pages} {093022} (\bibinfo {year} {2021})}\BibitemShut {NoStop}%
\bibitem [{\citenamefont {DeSalvo}\ \emph {et~al.}(2019)\citenamefont
  {DeSalvo}, \citenamefont {Patel}, \citenamefont {Cai},\ and\ \citenamefont
  {Chin}}]{DeSalvo2019Nature568.61--64}%
  \BibitemOpen
  \bibfield  {author} {\bibinfo {author} {\bibfnamefont {B.~J.}\ \bibnamefont
  {DeSalvo}}, \bibinfo {author} {\bibfnamefont {K.}~\bibnamefont {Patel}},
  \bibinfo {author} {\bibfnamefont {G.}~\bibnamefont {Cai}},\ and\ \bibinfo
  {author} {\bibfnamefont {C.}~\bibnamefont {Chin}},\ }\bibfield  {title}
  {\bibinfo {title} {Observation of fermion-mediated interactions between
  bosonic atoms},\ }\href {https://doi.org/10.1038/s41586-019-1055-0}
  {\bibfield  {journal} {\bibinfo  {journal} {Nature}\ }\textbf {\bibinfo
  {volume} {568}},\ \bibinfo {pages} {61} (\bibinfo {year} {2019})}\BibitemShut
  {NoStop}%
\bibitem [{\citenamefont {Edri}\ \emph {et~al.}(2020)\citenamefont {Edri},
  \citenamefont {Raz}, \citenamefont {Matzliah}, \citenamefont {Davidson},\
  and\ \citenamefont {Ozeri}}]{PhysRevLett.124.163401}%
  \BibitemOpen
  \bibfield  {author} {\bibinfo {author} {\bibfnamefont {H.}~\bibnamefont
  {Edri}}, \bibinfo {author} {\bibfnamefont {B.}~\bibnamefont {Raz}}, \bibinfo
  {author} {\bibfnamefont {N.}~\bibnamefont {Matzliah}}, \bibinfo {author}
  {\bibfnamefont {N.}~\bibnamefont {Davidson}},\ and\ \bibinfo {author}
  {\bibfnamefont {R.}~\bibnamefont {Ozeri}},\ }\bibfield  {title} {\bibinfo
  {title} {Observation of spin-spin fermion-mediated interactions between
  ultracold bosons},\ }\href {https://doi.org/10.1103/PhysRevLett.124.163401}
  {\bibfield  {journal} {\bibinfo  {journal} {Phys. Rev. Lett.}\ }\textbf
  {\bibinfo {volume} {124}},\ \bibinfo {pages} {163401} (\bibinfo {year}
  {2020})}\BibitemShut {NoStop}%
\bibitem [{\citenamefont {Baroni}\ \emph {et~al.}(2023)\citenamefont {Baroni},
  \citenamefont {Huang}, \citenamefont {Fritsche}, \citenamefont {Dobler},
  \citenamefont {Anich}, \citenamefont {Kirilov}, \citenamefont {Grimm},
  \citenamefont {Bastarrachea-Magnani}, \citenamefont {Massignan},\ and\
  \citenamefont {Bruun}}]{baroni2023mediated}%
  \BibitemOpen
  \bibfield  {author} {\bibinfo {author} {\bibfnamefont {C.}~\bibnamefont
  {Baroni}}, \bibinfo {author} {\bibfnamefont {B.}~\bibnamefont {Huang}},
  \bibinfo {author} {\bibfnamefont {I.}~\bibnamefont {Fritsche}}, \bibinfo
  {author} {\bibfnamefont {E.}~\bibnamefont {Dobler}}, \bibinfo {author}
  {\bibfnamefont {G.}~\bibnamefont {Anich}}, \bibinfo {author} {\bibfnamefont
  {E.}~\bibnamefont {Kirilov}}, \bibinfo {author} {\bibfnamefont
  {R.}~\bibnamefont {Grimm}}, \bibinfo {author} {\bibfnamefont {M.~A.}\
  \bibnamefont {Bastarrachea-Magnani}}, \bibinfo {author} {\bibfnamefont
  {P.}~\bibnamefont {Massignan}},\ and\ \bibinfo {author} {\bibfnamefont
  {G.}~\bibnamefont {Bruun}},\ }\bibfield  {title} {\bibinfo {title} {Mediated
  interactions between fermi polarons and the role of impurity quantum
  statistics},\ }\href {https://doi.org/10.1038/s41567-023-02248-4} {\bibfield
  {journal} {\bibinfo  {journal} {Nat. Phys.}\ }\textbf {\bibinfo {volume}
  {20}},\ \bibinfo {pages} {68} (\bibinfo {year} {2023})}\BibitemShut {NoStop}%
\bibitem [{\citenamefont {Ruderman}\ and\ \citenamefont
  {Kittel}(1954)}]{Ruderman1954Phys.Rev.96.99--102}%
  \BibitemOpen
  \bibfield  {author} {\bibinfo {author} {\bibfnamefont {M.~A.}\ \bibnamefont
  {Ruderman}}\ and\ \bibinfo {author} {\bibfnamefont {C.}~\bibnamefont
  {Kittel}},\ }\bibfield  {title} {\bibinfo {title} {Indirect exchange coupling
  of nuclear magnetic moments by conduction electrons},\ }\href
  {https://doi.org/10.1103/PhysRev.96.99} {\bibfield  {journal} {\bibinfo
  {journal} {Phys. Rev.}\ }\textbf {\bibinfo {volume} {96}},\ \bibinfo {pages}
  {99} (\bibinfo {year} {1954})}\BibitemShut {NoStop}%
\bibitem [{\citenamefont {Kasuya}(1956)}]{Kasuya1956Prog.Theor.Phys.16.45--57}%
  \BibitemOpen
  \bibfield  {author} {\bibinfo {author} {\bibfnamefont {T.}~\bibnamefont
  {Kasuya}},\ }\bibfield  {title} {\bibinfo {title} {{A Theory of Metallic
  Ferro- and Antiferromagnetism on Zener's Model}},\ }\href
  {https://doi.org/10.1143/PTP.16.45} {\bibfield  {journal} {\bibinfo
  {journal} {Prog. Theor. Phys.}\ }\textbf {\bibinfo {volume} {16}},\ \bibinfo
  {pages} {45} (\bibinfo {year} {1956})}\BibitemShut {NoStop}%
\bibitem [{\citenamefont {Yosida}(1957)}]{Yosida1957Phys.Rev.106.893--898}%
  \BibitemOpen
  \bibfield  {author} {\bibinfo {author} {\bibfnamefont {K.}~\bibnamefont
  {Yosida}},\ }\bibfield  {title} {\bibinfo {title} {Magnetic properties of
  cu-mn alloys},\ }\href {https://doi.org/10.1103/PhysRev.106.893} {\bibfield
  {journal} {\bibinfo  {journal} {Phys. Rev.}\ }\textbf {\bibinfo {volume}
  {106}},\ \bibinfo {pages} {893} (\bibinfo {year} {1957})}\BibitemShut
  {NoStop}%
\bibitem [{\citenamefont {Chen}\ \emph {et~al.}(2022)\citenamefont {Chen},
  \citenamefont {Duda}, \citenamefont {Schindewolf}, \citenamefont {Bause},
  \citenamefont {Bloch},\ and\ \citenamefont
  {Luo}}]{Chen2022Phys.Rev.Lett.128.153401}%
  \BibitemOpen
  \bibfield  {author} {\bibinfo {author} {\bibfnamefont {X.-Y.}\ \bibnamefont
  {Chen}}, \bibinfo {author} {\bibfnamefont {M.}~\bibnamefont {Duda}}, \bibinfo
  {author} {\bibfnamefont {A.}~\bibnamefont {Schindewolf}}, \bibinfo {author}
  {\bibfnamefont {R.}~\bibnamefont {Bause}}, \bibinfo {author} {\bibfnamefont
  {I.}~\bibnamefont {Bloch}},\ and\ \bibinfo {author} {\bibfnamefont {X.-Y.}\
  \bibnamefont {Luo}},\ }\bibfield  {title} {\bibinfo {title} {Suppression of
  unitary three-body loss in a degenerate bose-fermi mixture},\ }\href
  {https://doi.org/10.1103/PhysRevLett.128.153401} {\bibfield  {journal}
  {\bibinfo  {journal} {Phys. Rev. Lett.}\ }\textbf {\bibinfo {volume} {128}},\
  \bibinfo {pages} {153401} (\bibinfo {year} {2022})}\BibitemShut {NoStop}%
\bibitem [{\citenamefont {Naidon}\ and\ \citenamefont
  {Endo}(2017)}]{Naidon2017Rep.Prog.Phys.80.056001}%
  \BibitemOpen
  \bibfield  {author} {\bibinfo {author} {\bibfnamefont {P.}~\bibnamefont
  {Naidon}}\ and\ \bibinfo {author} {\bibfnamefont {S.}~\bibnamefont {Endo}},\
  }\bibfield  {title} {\bibinfo {title} {Efimov physics: a review},\ }\href
  {https://doi.org/10.1088/1361-6633/aa50e8} {\bibfield  {journal} {\bibinfo
  {journal} {Rep. Prog. Phys.}\ }\textbf {\bibinfo {volume} {80}},\ \bibinfo
  {pages} {056001} (\bibinfo {year} {2017})}\BibitemShut {NoStop}%
\bibitem [{\citenamefont {Tajima}\ \emph
  {et~al.}(2021{\natexlab{b}})\citenamefont {Tajima}, \citenamefont
  {Takahashi}, \citenamefont {Mistakidis}, \citenamefont {Nakano},\ and\
  \citenamefont {Iida}}]{Tajima2021Atoms9.18}%
  \BibitemOpen
  \bibfield  {author} {\bibinfo {author} {\bibfnamefont {H.}~\bibnamefont
  {Tajima}}, \bibinfo {author} {\bibfnamefont {J.}~\bibnamefont {Takahashi}},
  \bibinfo {author} {\bibfnamefont {S.~I.}\ \bibnamefont {Mistakidis}},
  \bibinfo {author} {\bibfnamefont {E.}~\bibnamefont {Nakano}},\ and\ \bibinfo
  {author} {\bibfnamefont {K.}~\bibnamefont {Iida}},\ }\bibfield  {title}
  {\bibinfo {title} {Polaron problems in ultracold atoms: Role of a fermi sea
  across different spatial dimensions and quantum fluctuations of a bose
  medium},\ }\href {https://doi.org/10.3390/atoms9010018} {\bibfield  {journal}
  {\bibinfo  {journal} {Atoms}\ }\textbf {\bibinfo {volume} {9}},\ \bibinfo
  {pages} {18} (\bibinfo {year} {2021}{\natexlab{b}})}\BibitemShut {NoStop}%
\bibitem [{\citenamefont {Patton}\ and\ \citenamefont
  {Sheehy}(2011)}]{Patton2011Phys.Rev.A83.051607}%
  \BibitemOpen
  \bibfield  {author} {\bibinfo {author} {\bibfnamefont {K.~R.}\ \bibnamefont
  {Patton}}\ and\ \bibinfo {author} {\bibfnamefont {D.~E.}\ \bibnamefont
  {Sheehy}},\ }\bibfield  {title} {\bibinfo {title} {Induced $p$-wave
  superfluidity in strongly interacting imbalanced fermi gases},\ }\href
  {https://doi.org/10.1103/PhysRevA.83.051607} {\bibfield  {journal} {\bibinfo
  {journal} {Phys. Rev. A}\ }\textbf {\bibinfo {volume} {83}},\ \bibinfo
  {pages} {051607} (\bibinfo {year} {2011})}\BibitemShut {NoStop}%
\bibitem [{\citenamefont {Drut}\ \emph {et~al.}(2018)\citenamefont {Drut},
  \citenamefont {McKenney}, \citenamefont {Daza}, \citenamefont {Lin},\ and\
  \citenamefont {Ord\'o\~nez}}]{Drut2018PhysRevLett.120.243002}%
  \BibitemOpen
  \bibfield  {author} {\bibinfo {author} {\bibfnamefont {J.~E.}\ \bibnamefont
  {Drut}}, \bibinfo {author} {\bibfnamefont {J.~R.}\ \bibnamefont {McKenney}},
  \bibinfo {author} {\bibfnamefont {W.~S.}\ \bibnamefont {Daza}}, \bibinfo
  {author} {\bibfnamefont {C.~L.}\ \bibnamefont {Lin}},\ and\ \bibinfo {author}
  {\bibfnamefont {C.~R.}\ \bibnamefont {Ord\'o\~nez}},\ }\bibfield  {title}
  {\bibinfo {title} {Quantum anomaly and thermodynamics of one-dimensional
  fermions with three-body interactions},\ }\href
  {https://doi.org/10.1103/PhysRevLett.120.243002} {\bibfield  {journal}
  {\bibinfo  {journal} {Phys. Rev. Lett.}\ }\textbf {\bibinfo {volume} {120}},\
  \bibinfo {pages} {243002} (\bibinfo {year} {2018})}\BibitemShut {NoStop}%
\bibitem [{\citenamefont {Guijarro}\ \emph {et~al.}(2018)\citenamefont
  {Guijarro}, \citenamefont {Pricoupenko}, \citenamefont {Astrakharchik},
  \citenamefont {Boronat},\ and\ \citenamefont {Petrov}}]{PhysRevA.97.061605}%
  \BibitemOpen
  \bibfield  {author} {\bibinfo {author} {\bibfnamefont {G.}~\bibnamefont
  {Guijarro}}, \bibinfo {author} {\bibfnamefont {A.}~\bibnamefont
  {Pricoupenko}}, \bibinfo {author} {\bibfnamefont {G.~E.}\ \bibnamefont
  {Astrakharchik}}, \bibinfo {author} {\bibfnamefont {J.}~\bibnamefont
  {Boronat}},\ and\ \bibinfo {author} {\bibfnamefont {D.~S.}\ \bibnamefont
  {Petrov}},\ }\bibfield  {title} {\bibinfo {title} {One-dimensional
  three-boson problem with two- and three-body interactions},\ }\href
  {https://doi.org/10.1103/PhysRevA.97.061605} {\bibfield  {journal} {\bibinfo
  {journal} {Phys. Rev. A}\ }\textbf {\bibinfo {volume} {97}},\ \bibinfo
  {pages} {061605} (\bibinfo {year} {2018})}\BibitemShut {NoStop}%
\bibitem [{\citenamefont {Sekino}\ and\ \citenamefont
  {Nishida}(2021)}]{Sekino2021Phys.Rev.A103.043307}%
  \BibitemOpen
  \bibfield  {author} {\bibinfo {author} {\bibfnamefont {Y.}~\bibnamefont
  {Sekino}}\ and\ \bibinfo {author} {\bibfnamefont {Y.}~\bibnamefont
  {Nishida}},\ }\bibfield  {title} {\bibinfo {title} {Field-theoretical aspects
  of one-dimensional bose and fermi gases with contact interactions},\ }\href
  {https://doi.org/10.1103/PhysRevA.103.043307} {\bibfield  {journal} {\bibinfo
   {journal} {Phys. Rev. A}\ }\textbf {\bibinfo {volume} {103}},\ \bibinfo
  {pages} {043307} (\bibinfo {year} {2021})}\BibitemShut {NoStop}%
\bibitem [{\citenamefont {Valiente}(2020)}]{PhysRevA.102.053304}%
  \BibitemOpen
  \bibfield  {author} {\bibinfo {author} {\bibfnamefont {M.}~\bibnamefont
  {Valiente}},\ }\bibfield  {title} {\bibinfo {title} {Bose-fermi dualities for
  arbitrary one-dimensional quantum systems in the universal low-energy
  regime},\ }\href {https://doi.org/10.1103/PhysRevA.102.053304} {\bibfield
  {journal} {\bibinfo  {journal} {Phys. Rev. A}\ }\textbf {\bibinfo {volume}
  {102}},\ \bibinfo {pages} {053304} (\bibinfo {year} {2020})}\BibitemShut
  {NoStop}%
\bibitem [{\citenamefont {Mistakidis}\ \emph {et~al.}(2023)\citenamefont
  {Mistakidis}, \citenamefont {Volosniev}, \citenamefont {Barfknecht},
  \citenamefont {Fogarty}, \citenamefont {Busch}, \citenamefont {Foerster},
  \citenamefont {Schmelcher},\ and\ \citenamefont {Zinner}}]{MISTAKIDIS20231}%
  \BibitemOpen
  \bibfield  {author} {\bibinfo {author} {\bibfnamefont {S.}~\bibnamefont
  {Mistakidis}}, \bibinfo {author} {\bibfnamefont {A.}~\bibnamefont
  {Volosniev}}, \bibinfo {author} {\bibfnamefont {R.}~\bibnamefont
  {Barfknecht}}, \bibinfo {author} {\bibfnamefont {T.}~\bibnamefont {Fogarty}},
  \bibinfo {author} {\bibfnamefont {T.}~\bibnamefont {Busch}}, \bibinfo
  {author} {\bibfnamefont {A.}~\bibnamefont {Foerster}}, \bibinfo {author}
  {\bibfnamefont {P.}~\bibnamefont {Schmelcher}},\ and\ \bibinfo {author}
  {\bibfnamefont {N.}~\bibnamefont {Zinner}},\ }\bibfield  {title} {\bibinfo
  {title} {Few-body bose gases in low dimensions—a laboratory for quantum
  dynamics},\ }\href
  {https://doi.org/https://doi.org/10.1016/j.physrep.2023.10.004} {\bibfield
  {journal} {\bibinfo  {journal} {Physics Reports}\ }\textbf {\bibinfo {volume}
  {1042}},\ \bibinfo {pages} {1} (\bibinfo {year} {2023})},\ \bibinfo {note}
  {few-body Bose gases in low dimensions—A laboratory for quantum
  dynamics}\BibitemShut {NoStop}%
\bibitem [{\citenamefont {McGuire}(1966)}]{McGuire1966J.Math.Phys.7.123--132}%
  \BibitemOpen
  \bibfield  {author} {\bibinfo {author} {\bibfnamefont {J.~B.}\ \bibnamefont
  {McGuire}},\ }\bibfield  {title} {\bibinfo {title} {{Interacting Fermions in
  One Dimension. II. Attractive Potential}},\ }\href
  {https://doi.org/10.1063/1.1704798} {\bibfield  {journal} {\bibinfo
  {journal} {J. Math. Phys.}\ }\textbf {\bibinfo {volume} {7}},\ \bibinfo
  {pages} {123} (\bibinfo {year} {1966})},\ \Eprint
  {https://arxiv.org/abs/https://pubs.aip.org/aip/jmp/article-pdf/7/1/123/10969537/123\_1\_online.pdf}
  {https://pubs.aip.org/aip/jmp/article-pdf/7/1/123/10969537/123\_1\_online.pdf}
  \BibitemShut {NoStop}%
\bibitem [{\citenamefont {Mao}\ \emph {et~al.}(2016)\citenamefont {Mao},
  \citenamefont {Guan},\ and\ \citenamefont {Wu}}]{Mao2016Phys.Rev.A94.043645}%
  \BibitemOpen
  \bibfield  {author} {\bibinfo {author} {\bibfnamefont {R.}~\bibnamefont
  {Mao}}, \bibinfo {author} {\bibfnamefont {X.~W.}\ \bibnamefont {Guan}},\ and\
  \bibinfo {author} {\bibfnamefont {B.}~\bibnamefont {Wu}},\ }\bibfield
  {title} {\bibinfo {title} {Exact results for polaron and molecule in
  one-dimensional spin-1/2 fermi gas},\ }\href
  {https://doi.org/10.1103/PhysRevA.94.043645} {\bibfield  {journal} {\bibinfo
  {journal} {Phys. Rev. A}\ }\textbf {\bibinfo {volume} {94}},\ \bibinfo
  {pages} {043645} (\bibinfo {year} {2016})}\BibitemShut {NoStop}%
\bibitem [{\citenamefont {Phyu}\ \emph {et~al.}(2020)\citenamefont {Phyu},
  \citenamefont {Moriya}, \citenamefont {Horiuchi}, \citenamefont {Iida},
  \citenamefont {Noda},\ and\ \citenamefont
  {Yamashita}}]{Phyu2020Prog.Theor.Exp.Phys.2020.093D01}%
  \BibitemOpen
  \bibfield  {author} {\bibinfo {author} {\bibfnamefont {L.~H.}\ \bibnamefont
  {Phyu}}, \bibinfo {author} {\bibfnamefont {H.}~\bibnamefont {Moriya}},
  \bibinfo {author} {\bibfnamefont {W.}~\bibnamefont {Horiuchi}}, \bibinfo
  {author} {\bibfnamefont {K.}~\bibnamefont {Iida}}, \bibinfo {author}
  {\bibfnamefont {K.}~\bibnamefont {Noda}},\ and\ \bibinfo {author}
  {\bibfnamefont {M.~T.}\ \bibnamefont {Yamashita}},\ }\bibfield  {title}
  {\bibinfo {title} {{Coulomb screening correction to the Q value of the
  triple-alpha process in thermal plasmas}},\ }\href
  {https://doi.org/10.1093/ptep/ptaa093} {\bibfield  {journal} {\bibinfo
  {journal} {Prog. Theor. Exp. Phys.}\ }\textbf {\bibinfo {volume} {2020}},\
  \bibinfo {pages} {093D01} (\bibinfo {year} {2020})},\ \Eprint
  {https://arxiv.org/abs/https://academic.oup.com/ptep/article-pdf/2020/9/093D01/33701430/ptaa093.pdf}
  {https://academic.oup.com/ptep/article-pdf/2020/9/093D01/33701430/ptaa093.pdf}
  \BibitemShut {NoStop}%
\bibitem [{\citenamefont {McGuire}(1964)}]{McGuire2004J.Math.Phys.5.622--636}%
  \BibitemOpen
  \bibfield  {author} {\bibinfo {author} {\bibfnamefont {J.~B.}\ \bibnamefont
  {McGuire}},\ }\bibfield  {title} {\bibinfo {title} {{Study of Exactly Soluble
  One‐Dimensional N‐Body Problems}},\ }\href
  {https://doi.org/10.1063/1.1704156} {\bibfield  {journal} {\bibinfo
  {journal} {J. Math. Phys.}\ }\textbf {\bibinfo {volume} {5}},\ \bibinfo
  {pages} {622} (\bibinfo {year} {1964})},\ \Eprint
  {https://arxiv.org/abs/https://pubs.aip.org/aip/jmp/article-pdf/5/5/622/8174007/622\_1\_online.pdf}
  {https://pubs.aip.org/aip/jmp/article-pdf/5/5/622/8174007/622\_1\_online.pdf}
  \BibitemShut {NoStop}%
\bibitem [{\citenamefont {Braaten}\ \emph {et~al.}(2017)\citenamefont
  {Braaten}, \citenamefont {Hammer},\ and\ \citenamefont
  {Lepage}}]{Braaten2017Phys.Rev.A95.012708}%
  \BibitemOpen
  \bibfield  {author} {\bibinfo {author} {\bibfnamefont {E.}~\bibnamefont
  {Braaten}}, \bibinfo {author} {\bibfnamefont {H.-W.}\ \bibnamefont
  {Hammer}},\ and\ \bibinfo {author} {\bibfnamefont {G.~P.}\ \bibnamefont
  {Lepage}},\ }\bibfield  {title} {\bibinfo {title} {Lindblad equation for the
  inelastic loss of ultracold atoms},\ }\href
  {https://doi.org/10.1103/PhysRevA.95.012708} {\bibfield  {journal} {\bibinfo
  {journal} {Phys. Rev. A}\ }\textbf {\bibinfo {volume} {95}},\ \bibinfo
  {pages} {012708} (\bibinfo {year} {2017})}\BibitemShut {NoStop}%
\bibitem [{\citenamefont {Sun}\ \emph {et~al.}(2017)\citenamefont {Sun},
  \citenamefont {Zhai},\ and\ \citenamefont
  {Cui}}]{Sun2017Phys.Rev.Lett.119.013401}%
  \BibitemOpen
  \bibfield  {author} {\bibinfo {author} {\bibfnamefont {M.}~\bibnamefont
  {Sun}}, \bibinfo {author} {\bibfnamefont {H.}~\bibnamefont {Zhai}},\ and\
  \bibinfo {author} {\bibfnamefont {X.}~\bibnamefont {Cui}},\ }\bibfield
  {title} {\bibinfo {title} {Visualizing the efimov correlation in bose
  polarons},\ }\href {https://doi.org/10.1103/PhysRevLett.119.013401}
  {\bibfield  {journal} {\bibinfo  {journal} {Phys. Rev. Lett.}\ }\textbf
  {\bibinfo {volume} {119}},\ \bibinfo {pages} {013401} (\bibinfo {year}
  {2017})}\BibitemShut {NoStop}%
\bibitem [{\citenamefont {Huber}\ \emph {et~al.}(2019)\citenamefont {Huber},
  \citenamefont {Hammer},\ and\ \citenamefont
  {Volosniev}}]{PhysRevResearch.1.033177}%
  \BibitemOpen
  \bibfield  {author} {\bibinfo {author} {\bibfnamefont {D.}~\bibnamefont
  {Huber}}, \bibinfo {author} {\bibfnamefont {H.-W.}\ \bibnamefont {Hammer}},\
  and\ \bibinfo {author} {\bibfnamefont {A.~G.}\ \bibnamefont {Volosniev}},\
  }\bibfield  {title} {\bibinfo {title} {In-medium bound states of two bosonic
  impurities in a one-dimensional fermi gas},\ }\href
  {https://doi.org/10.1103/PhysRevResearch.1.033177} {\bibfield  {journal}
  {\bibinfo  {journal} {Phys. Rev. Res.}\ }\textbf {\bibinfo {volume} {1}},\
  \bibinfo {pages} {033177} (\bibinfo {year} {2019})}\BibitemShut {NoStop}%
\end{thebibliography}


\end{CJK}
\end{document}